\documentclass[aps,prd,superscriptaddress,nofootinbib,amsmath,amsfonts,preprintnumbers,groupedaddress,showpacs,10pt,english]{revtex4-1}
\usepackage{amsmath}
\usepackage{amssymb}
\usepackage{babel}
\usepackage{wrapfig}
\usepackage{cancel}

\newcommand{\e}{\mathrm{e}}
\usepackage{relsize,exscale}
\makeatletter

\usepackage{array,multirow,graphicx}
\usepackage{dcolumn}
\usepackage{newlfont}
\usepackage{bm}
\usepackage[colorlinks,citecolor=blue,urlcolor=blue,linkcolor=blue]{hyperref}
\usepackage[figtopcap]{subfigure}
\usepackage{color}

\begin{document}
\tolerance=5000
\title{Charged  solution with equal metric ansatz in Gauss-Bonnet theory coupled to scalar field}

\author{G.~G.~L.~Nashed$^{1,2}$}\email{nashed@bue.edu.eg}

\affiliation {$^1$ Centre for Theoretical Physics, The British University, P.O. Box
43, El Sherouk City, Cairo 11837, Egypt \\
$^2$Center for Space Research, North-West University, Potchefstroom 2520, South Africa\\
}
\date{}
\begin{abstract}
In the course of this research, we employ the Gauss-Bonnet equation of motion alongside the scalar field and potential to acquire a fresh solution for a spherically symmetrical charged black hole. Specifically, we derive this black hole solution by employing a metric potential where the components are equal, that is, $g_{tt}=g_{rr}$. In our research, we achieve several accomplishments, including fixing the characteristics of the scalar field, and the Gauss-Bonnet term. We thoroughly examine the physical properties associated with such black hole   and show that we have supplementary terms when compared to the Reissner-Nordström black hole solution. These additional terms are of the order $O(\frac{1}{r^6})$ and $O(\frac{1}{r^9})$. The presence of these supplementary terms can be attributed to the impact of the scalar function denoted as $\xi$. Such expressions play a crucial role in generating the multi-horizon black hole solution. The presence of these extra terms facilitates the derivation of a modified first law of thermodynamics and the corresponding Smarr relation.
\end{abstract}


\maketitle
\section{Introduction}\label{S1}
Currently, black holes (BHs), which were previously seen as a matter of mathematical fascination within Einstein's general relativity (GR)  theory, have been discovered to proliferate in significant quantities throughout the vastness of our universe. At present, BHs are observed to fall into two distinct classifications: stellar black holes, which have masses ranging from 5 to 70 times that of the Sun ($M_\odot$), and supermassive BHs located  at the core of galaxies, BHs with masses that extend to$10^{10},M_\odot$. A BH represents the most straightforward example of the behavior of gravity under extreme conditions, making it a valuable experimental tool for investigating the fundamental principles of gravitational forces.

While GR is an elegant mathematical framework that possesses successfully withstood numerous experimental validations (consider, for example,\cite{Bertotti:2003rm, Stairs:2003eg,Verma:2013ata,Titov:2010zn, Will:2014kxa, Kramer:2021jcw}), it is clear that it falls short in providing comprehensive explanations for numerous persistent questions in the realms of cosmology and gravity. Among the challenges that remain unsolved, some noteworthy ones include the presence  of singularities, the perplexing properties of  dark energy and dark matter, the obstacles encountered in the process of quantifying gravity, as well as the quest for unifying gravity with the other fundamental forces of nature. It is widely accepted within the scientific community that GR is merely an approximation at low energies and that there exists a more basic explanation of gravity yet to be discovered. Since the precise framework for a complete  understanding of gravity at the quantum level remains elusive, the prevailing approach in the meantime is to employ effective field theory. Within this approach, GR, which is a linear theory based on curvature, is augmented by incorporating higher-order gravitational terms, introducing extra fields (gauge fields and primarily scalar), and exploring new connections, like higher-order interactions, between gravity and matter, with the goal of encompassing the pertinent phenomena beyond classical physics and accommodating potential alterations to the nature of gravity.

Attempting to expand GR in this manner unavoidably leads to a significantly wider array of gravitational solutions. Specifically, within the framework of modified gravity theories, it has long been acknowledged that new solutions for black holes exist  \cite{Luckock:1986tr, Bizon:1990sr, Campbell:1990ai, Campbell:1991kz, Maeda:1993ap, Kanti:1995vq, Kanti:1996gs, Kanti:1997br, Torii:1996yi}. By circumventing the no-hair theorems of GR  (refer to \cite{Kerr:1963ud, Israel:1967wq, Israel:1967za, Carter:1968rr, Carter:1971zc, Hawking:1971vc, Price:1971fb, Price:1972pw, Robinson:1975bv, Teitelboim:1972qx, Bekenstein:1971hc, Bekenstein:1972ky, Bekenstein:1995un}), several additional solutions have emerged in recent years within modified gravity theories (see \cite{Guo:2008hf, Pani:2009wy, Pani:2011gy, Kleihaus:2011tg,Yagi:2012ya, Sotiriou:2013qea, Sotiriou:2014pfa, Kleihaus:2015aje, Blazquez-Salcedo:2016enn, Antoniou:2017acq, Antoniou:2017hxj, Doneva:2017bvd, Silva:2017uqg, Brihaye:2017wln, Doneva:2018rou, Bakopoulos:2018nui, Witek:2018dmd, Minamitsuji:2018xde, Macedo:2019sem, Doneva:2019vuh, Zou:2019bpt, Cunha:2019dwb, Brihaye:2019dck}). Moreover, these modified theories also predict compact solutions that deviate from traditional BHs, such as traversable wormholes (see \cite{Bronnikov:1973fh, Ellis:1973yv, Visser:2003yf, Bronnikov:1996de, Bronnikov:1997gj, Armendariz-Picon:2002gjc, Bronnikov:2004ax, Lobo:2005us, Lobo:2005vc, Lobo:2009ip, Bronnikov:2010tt, Garcia:2010xb, Kanti:2011jz, Kanti:2011yv, Bolokhov:2012kn, Bronnikov:2015pha, Shaikh:2015oha, Mehdizadeh:2015jra, Kuhfittig:2018vdg, Ibadov:2021oqf, Karakasis:2021tqx, Ghosh:2021dgm}), as well as particle-like solutions (see \cite{Fisher:1948yn, Janis:1968zz, Wyman:1981bd, Nashed:2011fg, Agnese:1985xj, Roberts:1989sk, Kleihaus:2019rbg, Kleihaus:2020qwo}). The concept of these condensed entities existing throughout the universe no longer appears implausible and has become an intriguing possibility.

This study focuses on a particular group of altered theories of gravity that include scalar function. The scalar function  can be generated through gravitational fields,resulting in solutions exhibiting scalarization either caused by external forces or occurring naturally, or via interactions with gauge fields, leading to charged scalarized solutions. Using the theory of Gauss Bonnet  alongside a scalar function to analyze a spherically symmetric black holes  provides numerous advantages, as it not only alters the gravitational backdrop but also impacts the geodesic structure of spacetime. This influence extends to the trajectories of photons and the size and arrangement of the black hole's silhouette. It enables the inclusion of adjustments for increased curvature, the prevention of singularities, the investigation of alternative theories of gravity, the verification of theoretical forecasts, and the establishment of links to fundamental theories in physics. In this inquiry, we aim to present a concise summary of the constraints related to theories that encompass the Gauss-Bonnet scalar. These theories hold substantial importance in the realm of string theories. To clarify further, our primary focus will revolve around the presence of spectral entities within these hypotheses. These spectral entities are frequently associated with gravitational theories that involve higher-order derivatives, as they manifest due to the instability referred to as Ostrogradsky's apparition \cite{Woodard:2015zca}. The issue of these additional unobservable variables is discussed in the paper by \cite{DeFelice:2009ak}, indicating their potential manifestation in diverse facets of the theory, including cosmological perturbations within $f(R, \mathcal{G})$ theories. The primary aim of this investigation is to discover solutions for spherically symmetric charged black holes within the gravitational framework of $f(\mathcal{G})$ theory, which does not incorporate these unobservable variables. This theory was initially introduced in.~\cite{Nojiri:2018ouv, Nojiri:2021mxf}.

 Our specific approach involves utilizing the  equation of motions of the Einstein-Gauss-Bonnet theory with scalar field, (ESGB),  in the context of charged systems. In a four-dimensional spacetime, the GB term has a topological property and does not have any substantial dynamic influence.  Nevertheless, when this expression is when coupled in a non-minimal fashion with a scalar field labeled as $\xi$, the resulting dynamics display non-trivial characteristics.Several cosmological investigations have been presented in recent scientific publications \cite{Brax:2003fv,Nojiri:2005jg,Nojiri:2005am,Cognola:2006eg,Nojiri:2010oco,Cognola:2009jx,Capozziello:2008gu,Sadeghi:2009pu,Guo:2010jr,Satoh:2010ep, Nozari:2013wua,Lahiri:2016qih,Mathew:2016anx,Nozari:2015jca,Motaharfar:2016dqt,Carter:2005fu,DeLaurentis:2015fea,vandeBruck:2016xvt,Granda:2014zea,Granda:2011kx,Nojiri:2005vv,Hikmawan:2015rze,Kanti:1998jd,Easther:1996yd,
 Rizos:1993rt,Starobinsky:1980te,Mukhanov:1991zn,Brandenberger:1993ef,Barrow:1993hp,Damour:1994zq,Angelantonj:1994dv,Kaloper:1995tu,Gasperini:1996in,
Rey:1996ad,Rey:1996ka,Easther:1995ba,Nashed:2021sji,Santillan:2017nik,Bose:1997qv,KalyanaRama:1996ar,KalyanaRama:1996im,KalyanaRama:1997xt,Brustein:1997ny,Brustein:1997cv} and references therein. The work introduced in the paper has derived a static black hole solution within the framework of Gauss-Bonnet gravity theory, in a spacetime with more than four dimensions \cite{Chen:2017hwm}.  The study presented in the paper has undertaken an examination of charged solutions within the context of the dilatonic Einstein-Gauss-Bonnet theory \cite{Melis:2005ji}. Moreover, the exploration of critical behavior related to asymptotically flat, charged, non-rotating solutions within a five-dimensional Einstein-Gauss-Bonnet gravity model, with minimal coupling to a U(1) charged scalar field exhibiting harmonic time-dependence, is investigated in the study presented in \cite{Brihaye:2018nta}.  To the best of our understanding, the particular instance of a charged spherically symmetric black hole solution within the Gauss-Bonnet theory, in combination with a non-minimal scalar field, potential, and an arbitrary scalar field function, has not been previously explored or examined in existing literature.  The aim of this study is to derive a precise solution for a charged spherically symmetric black hole within the context of the theory mentioned above and analyze its physical consequences. The methodology used in this research can be summarized as follows:

Section~\ref{GBS} provides an overview of the key aspects of the  gravitational theory that is free from ghosts, $f(\mathcal{G})$, which serves as the foundation for describing the formation of black hole horizons.

In Section~\ref{S2}, we employ the equations of motion for charged systems, as discussed in Section~\ref{S22}, to a spherically symmetric spacetime characterized by the condition $g_{tt}=g_{rr}$.

Section~\ref{S3} focuses on exploring the physics of the derived charged  solution, along with a discussion of relevant physics. Furthermore, in this section, we deduce the altered first law of thermodynamics and establish the corresponding Smarr relation.

The stability of this model is examined in Section~\ref{S4} using geodesic deviation analysis.

Section~\ref{S5} provides a summary of the main findings from this study and suggests potential avenues for future research.
\section{A concise overview of the $f(\mathcal{G})$   theory free ghosts }\label{S22}

In this part, we provide a concise explanation of how the  $f\left(\mathcal{G} \right)$ theory which is free from ghosts  is built utilizing the Lagrange multipliers.
Furthermore, we will explore the process of obtaining $f(\mathcal{G})$ theory that is free from ghosts and employ the technique of Lagrange multipliers for its formulations.
Prior to delving into the specifics of such formulation, we initiate the construction by examining the intricacies of the presence of ghosts in $f\left(\mathcal{G}\right)$ theory within the framework of the equation of motions. Subsequently, we proceed to formulate the ghost-free theory.

\subsection{Presence of spectral entities in $f\left( \mathit{G} \right)$ Gravitational Theory}\label{GBS}
For the readers that are interested in more details of the ESGB theory please see \cite{Nojiri:2018ouv, Nojiri:2021mxf} and references therein.

We start our study by writing the action of the theory under consideration as follows:
\begin{align}
\label{FRGBg222}
S=\int d^4x\sqrt{-g} \left(\frac{1}{2\kappa^2}R
+ \lambda \left( \frac{1}{2}\omega(\xi) \partial_\mu \xi \partial^\mu \xi + \frac{\mu^4}{2} \right)
+ h\left( \xi \right) \mathit{G} - \tilde V\left( \xi \right)
+ \mathcal{L}_\mathrm{matter}-\Lambda +\mathcal{L}_\mathrm{em}\right)\, ,
\end{align}
In this research, we express the Lagrangian for the electromagnetic field, denoted as $\mathcal{L}_\mathrm{em}$, as follows:
$\mathcal{L}_\mathrm{em}= - \frac{1}{2}\mathfrak{F}$
$\left( \mathfrak{F}\equiv \mathfrak{F}_{\mu \nu}\mathfrak{F}^{\mu \nu}\right)$ where
$\mathfrak{F} = d\zeta$ and
$\zeta=\zeta_{\beta}dx^\beta$ represents the 1-form electromagnetic potential.\cite{Capozziello:2012zj}.

The variations of the action, as described in Eq.~(\ref{FRGBg222}), concerning the metric, along with the auxiliary field $\xi$ and the Lagrange multiplier $\lambda$, results in the following:
\begin{align}
 \label{FRGBg23gg}
0 =& - \frac{1}{\sqrt{-g}} \partial_\mu \left( \lambda \omega(\xi) g^{\mu\nu}\sqrt{-g}
\partial_\nu \xi \right)
+ h'\left( \xi \right) \mathit{G} - {\tilde V}'\left( \xi \right) +\frac{1}{2} \lambda \omega'(\xi) g^{\mu\nu}
\partial_\mu \xi\partial_\nu \xi\, ,\\
\label{FRGBg23}
 0=&\frac{1}{2}\omega(\xi) \partial_\mu \xi \partial^\mu \xi + \frac{\mu^4}{2}\,,\\
\label{FRGBg24BB}
0 =& \frac{1}{2\kappa^2}\left(- R_{\mu\nu}
+ \frac{1}{2}g_{\mu\nu} [R-4\Lambda]\right) + \frac{1}{2} T_{\mathrm{matter}\, \mu\nu}
 - \frac{1}{2} \lambda \omega(\xi) \partial_\mu \xi \partial_\nu \xi
 - \frac{1}{2}g_{\mu\nu} \tilde V \left( \xi \right)
+ D_{\mu\nu}^{\ \ \tau\eta} \nabla_\tau \nabla_\eta h \left( \xi
\right)+T_{\mathrm{em}\, \mu\nu}\, .
\end{align}
In this investigation, the energy-momentum tensor, denoted as $T_{\mathrm{em}, \mu\nu}$, is derived from the Maxwell field as follows:
\begin{equation}
\label{en11}
{T_{\mathrm{em}\, \mu}}^\nu=\mathfrak{F}_{\mu \alpha}\mathfrak{F}^{\nu\alpha}
 -\frac{1}{4} \delta_\mu{}^\nu \mathfrak{F}\, .
\end{equation}
Furthermore, by taking the variation of Eq.~(\ref{FRGBg222}) with respect to the 1-form gauge potential, denoted by $\zeta_{\mu}$, we obtain:
\begin{equation}
\label{q8b}
L^\nu\equiv \partial_\mu\left( \sqrt{-g} \mathfrak{F}^{\mu \nu} \right)=0\, .
\end{equation}
In the upcoming sections, we do not consider the matter energy-momentum tensor, denoted as $\mathcal{L}\mathrm{matter}$, because our focus is on finding solutions in a vacuum. Instead, we utilize the field equations presented in equations (\ref{FRGBg23gg}), (\ref{FRGBg23}), (\ref{FRGBg24BB}), and \eqref{q8b}, which characterize a spherically symmetric spacetime with identical metric potentials, specifically $g_{tt}=g_{rr}$. We then examine the obtained solutions from a physics perspective.

\section{Spherically symmetric charged black hole solution in four dimensions}\label{S2}
Given the metric of the spherically symmetric spacetime in the following form:
\begin{equation}
\label{met}
ds^2 = b(r)dt^2 - \frac{dr^2}{b(r)} - r^2 \left[ d\theta^2 + \sin^2 \left( \theta \right) \right] d\phi^2\,,
\end{equation}
where $b(r)$ represents an unspecified function of the radial coordinate, $r$. It's important to highlight that the assumption of equal metric potentials may not be applicable in certain situations, such as when analyzing rotating black holes, which require additional terms for a comprehensive system description. Nonetheless, in spherically symmetric scenarios, adopting the assumption of equal metric potentials is a sensible and frequently utilized simplification that aligns with the fundamental principles of gravitational physics and is consistent with astrophysical observations.
Utilizing the field equations, as described in Eqs.~ (\ref{FRGBg23gg}), (\ref{FRGBg23}), (\ref{FRGBg24BB}), and \eqref{q8b}, within the spacetime framework outlined in Eq.~(\ref{met}), we obtain:\\
\\
The component of the field equation in the $(t,t)$ direction, as denoted by Eq.~ (\ref{FRGBg23gg}), can be expressed as:
\begin{align}
\label{eqtt}
&0=-\frac {1 }{{r}^{4} \sin^2\theta}\bigg\{12\, h'b  b'{ r}^{2} \sin^2\theta - b' {r}^{3}\sin^2 \theta-\tilde V{r}^{4} \sin^2 \theta+\Lambda\,{r}^{4} \sin^2\theta-8\,{r}^{2} \sin^2\theta h''b  +8\,{r}^{2} \sin^2\theta h''  b^{2}-4\, h' b' {r}^{2} \sin^2\theta\nonumber\\
&-b {r}^{2} \sin^2\theta+ k_\theta{}^{2}+{r}^{2} \sin^2\theta+ q'^{2}{r}^{4} \sin^2\theta+b {r}^{2} n_\phi^{2}+b {r}^{2} m'^{2}-2\,b  {r}^{2}n_\phi m'\bigg\} \,.
\end{align}
The component of the field equation in the $(t,\phi)$ direction, as denoted by Eq.~ (\ref{FRGBg23gg}), can be expressed as:
\begin{align}
\label{eqtphi}
0=-\frac{2(n_\phi-m')bq'}{r^2\sin^2\theta}\,.
\end{align}
The component of the field equation in the $(r,r)$ direction, as denoted by Eq.~ (\ref{FRGBg23gg}), can be expressed as:
\begin{align}
\label{eqrr}
&0=\frac {1}{{r}^{4} \sin^2\theta}\bigg\{b'{r}^{3} \sin^2 \theta-12\,h'b   b'{ r}^{2}\sin^2 \theta+\tilde V {r}^{4} \sin^2\theta-\Lambda\,{r}^{4} \sin^2\theta+4\,h' b'{r}^{2} \sin^2\theta+ \lambda \omega  \xi'^{2}b  {r}^{4} \sin^2\theta+b  {r }^{2} \sin^2\theta- k_\theta^{2}\nonumber\\
&-{r}^{2} \sin^2\theta- q'^{2}{r}^{4} \sin^2 \theta +b {r}^{2} n_\phi^{2}+b {r}^{2} m'^{2}-2\,b{ r}^{2} n_\phi m'\bigg\}\,.
\end{align}
The component of the field equation in the $(r,\phi)$ direction, as denoted by Eq.~ (\ref{FRGBg23gg}), can be expressed as:
\begin{align}
\label{eqrphi1}
0=-\frac{2(n_\phi-m')k_\theta}{r^4\sin^2\theta}\,.
\end{align}
The component of the field equation in the $(\theta,r)$ direction, as denoted by Eq.~ (\ref{FRGBg23gg}), can be expressed as:
\begin{align}
\label{eqrphi2}
0=-\frac{2bk_\theta(n_\phi-m')}{r^2\sin^2\theta}\,.
\end{align}
The component of the field equation in the $(\theta,\theta)$ direction, as denoted by Eq.~ (\ref{FRGBg23gg}), can be expressed as:
\begin{align}
\label{eqthth}
&0={\frac {1}{2{r}^{4} \sin^2\theta}}\bigg\{{r}^{3} \sin^2\theta \left( r-8\, h' b   \right) b''  -8\,{r}^{3} \sin^2\theta h'' b b'  -8\,{r}^{3} \sin^2\theta h'  b'^ {2}+2\,b' {r}^{3} \sin^2\theta-2\,b  {r}^{2}n_\phi^{2}+4\,b {r}^{2} n_\phi m' -2\,b  {r}^{2} m'^{2}\nonumber\\
&+2\,k_\theta^{2}-2\,{r}^{4} \sin^2\theta^{ 2} \left( \Lambda- q'^{2}-\tilde V \right)\bigg\}\,.
\end{align}
The component of the field equation in the $(t,\phi)$ direction, as denoted by Eq.~ (\ref{FRGBg23gg}), can be expressed as:
\begin{align}
\label{eqphi3}
0=2q'(n_\phi-m')\,,
\end{align}
and the component of the field equation in the $(\phi,\phi)$ direction, as denoted by Eq.~ (\ref{FRGBg23gg}), can be expressed as:
\begin{align}
&0={\frac {1}{2{r}^{4} \sin^2 \theta }}\bigg\{{r}^{3} \sin^2\theta\left(r -8\, h'b  \right) b'' -8\,{r}^{3} \sin^2\theta\,h'' b'b -8\,{r}^{3} \sin^2\theta h' b'^ {2}+2\, b'{r}^{3} \sin^2 \theta +2\,b {r}^{2} n_\phi^{2}-4\,b {r}^{2} n_\phi m'  +2\,b {r}^{2} m'^{2}\nonumber\\
&+2\, k_\theta^{2}-2\,{r}^{4} \sin^2\theta \left(\Lambda-\tilde V- q'^{2} \right)\bigg\}\,.
\end{align}
Here $k_\theta=\frac{\partial k(\theta)}{\partial \theta}$, $n_\phi=\frac{\partial n(\phi)}{\partial \phi}$, $q'=\frac{\partial q(r)}{\partial r}$, $m'=\frac{\partial m(r)}{\partial r}$.
The equations governing the scalar field, as described in Eqs.~(\ref{FRGBg23gg}) and \eqref{FRGBg23}, are represented in the following manner:
\begin{align}
\label{eqphi11}
0= \frac {1}{{r}^{2}\xi' }\bigg\{8\, h'  \left( b -1 \right) b'' -2\,{r}^{2}\xi' \lambda \omega b  \xi''  -2\,r \left\{ r\lambda \omega\,b'  + \left[ r \lambda' \omega +1/2\, \lambda \left( 4\,\omega +r\omega'   \right) \right] b  \right\}\xi'^ {2}-2\,V' {r}^{2}+8\, h'  b'^{2}\bigg\}
\,,
\end{align}
\begin{align}
\label{eqphi1}
0={\frac {r\lambda  \omega  b \xi''  + \left\{ r \lambda' \omega\,b + \left[ r\omega'  b  +\omega \left( b'  r+2\,b  \right)  \right] \lambda   \right\} \xi'  }{r}}\,.
\end{align}
Ultimately, the non-zero elements of the field equations \eqref{q8b} result in the following expression:
\begin{align}
\label{eqphie}
L^t\equiv-\frac{q''+2q'}{r}=0\,, \qquad L^r\equiv\frac{b\,n_{\phi \phi}}{r^2\sin^2\theta}=0\,, \qquad L^\phi\equiv-\frac{r^2\sin\theta(b'[n_\phi-m']-b\,m'')+k_\theta\cos\theta -k_{\theta\theta}\sin\theta}{r^4\sin^3\theta}=0\,,
\end{align}
where $'=\frac{d}{dr}$, $''=\frac{d^2}{dr^2}$, $n_\phi=\frac{dn(\phi)}{d\phi}$, and $k_\theta=\frac{dk(\theta)}{d\theta}$.

During this study, we examine the Maxwell field's vector potential in the given format:
\begin{align}
\label{vect} \zeta=q(r)dt+n(\phi)dr+[m(r)+k(\theta)]d\phi\,.
\end{align}
Here, $q(r)$ and $m(r)$ vary with the radial coordinate, $n(\phi)$ depends on $\phi$, and $k(\theta)$ depends on $\theta$. The solution for the non-diagonal components, as presented in Eqs. \eqref{eqtphi}, \eqref{eqrphi1}, \eqref{eqrphi2}, and \eqref{eqphi3}, can be expressed as follows:
\begin{align}
n_\phi=m'\,.\end{align}
Utilizing the second equation from Eq. \eqref{eqphie}, we obtain:
\begin{align}
\label{sol1}
n(\phi)=c_0\phi\,, \qquad {\textrm and} \qquad m(r)=c_0\,r\,.\end{align}
Similarly, the initial equation in Eq. \eqref{eqphie} provides:
\begin{align}
\label{sol2}
q(r)=\frac{c_1}{r}\,.\end{align}
Applying Eq.~(\ref{sol1}) and \eqref{sol2} to the third equation in Eq. \eqref{eqphie} results in:
\begin{align}
\label{sol3}
k(\theta)=c_2\cos\theta\,.\end{align}
Furthermore, for simplifying the computations, we will suppose that the scalar field can be represented as \begin{align}\xi=c_3,r.\end{align} Substituting this information into Eqs. \eqref{eqtt}, \eqref{eqrr}, \eqref{eqthth}, \eqref{eqphi11}, and \eqref{eqphi1}, we obtain:
\begin{align}
\label{sol4}
b(r)=\Lambda r^2+1+\frac{c_4}{r}+\frac{c_1+c_2}{r^2}+\frac{c_3}{r^6}+\frac{c_3{}^{3/2}}{r^9}\,.
\end{align}
Given the extensive expressions for the variables $\lambda$, $h(r)$, ${\tilde V}$, we have included them in an Appendix. The details of $\omega$ are also provided in the appendix.

\section{BH Physical behavior}\label{S3}
In this part, we pull out the physics of the BH as determined in the preceding section. To achieve this goal, we will be writing the line element of this BH in the following form:
\begin{align}\label{met11}
&ds^2=\bigg[\Lambda r^2+1-\frac{2M}{r}+\frac{q_e+q_m}{r^2}+\frac{c_3}{r^6}+\frac{c_3{}^{3/2}}{r^9}\bigg]dt^2-\frac{dr^2}{\Lambda r^2+1-\frac{2M}{r}+\frac{q_e+q_m}{r^2}+\frac{c_3}{r^6}+\frac{c_3{}^{3/2}}{r^9}}-r^2(d\theta^2+\sin^2\theta d\phi^2)\,,
\end{align}
where we have put $c_4=-2M$, $c_1=q_e$ and $c_2=q_m$. The metric above indicates that it is achievable to get the Reissner-Nordsr\"om  BH when the dimensional constant  $c_3$ vanishes. { This dimensional constant is not related to the electric nor the magnetic fields however,  its contribution comes from the scalar field, $\xi=c_3\,r$, as well as the GB term as shown in Eq. (\ref{inv}).}
 The solution presented in Eq.~\eqref{met11} for a  BH is distinguished by three main components: the mass denoted as $M$, the dimensional parameters $q_e$ and $q_m$ that correspond to the electric and magnetic fields respectively, and the dimensional parameter $c_3$. It is important to emphasize that when the values of $q_m$ and $c_3$ are both set to zero, the solution reverts back to the Reissner-Nordstr{" o}m BH solution.   However, if we assign zero values to the dimensional constants $c_3$, $q_e$, and $q_m$, we can obtain the Schwarzschild black hole solution, which corresponds to Einstein's general relativity. When we take the limit of $c_3=0$ and $q_e=q_m=0$, the scalar field and Lagrangian multiplier, denoted as $\lambda$ and $\xi$ respectively, both become zero. Even though $c_3=0$ and $q_e=q_m=0$, the arbitrary function $h$ remains non-zero. However, this does not result in a new black hole distinct from the Schwarzschild spacetime because the Schwarzschild solution already has a vanishing Gauss-Bonnet term.
 The scalars associated with the solution given by Eq. (\ref{met11}) take the form:
 \begin{align}
 \label{inv}
 &R^{\alpha\beta \gamma \rho}R_{\alpha\beta \gamma \rho}=24\Lambda^2+\frac{48M^2}{r^6}-\frac{96M(q_e+q_m)}{r^7}+\frac{80Mc_3+56(q_e{}^2+q_m{}^2)}{r^8}
 \,,\nonumber\\
 &R^{\alpha\beta}R_{\alpha\beta}=36\Lambda^2+\frac{120\Lambda c_3+4(q_e{}^2+q_m{}^2)}{r^8}
 \,,\nonumber\\
 &R=-12\Lambda-\frac{20c_3}{r^8}
 \nonumber\\
 &G=24\Lambda^2+\frac{48M^2}{r^6}-\frac{96M(q_e+q_m)}{r^7}+\frac{80Mc_3+40(q_e{}^2+q_m{}^2)}{r^8}
 \,.
\end{align}
Eq.~(\ref{inv}) shows that the above invariant has  constant values as $r\to \infty$ and becomes infinite as $r\to 0$.

\begin{figure}
\centering
\subfigure[~The metric given by Eq.~(\ref{met11})]{\label{fig:met}\includegraphics[scale=0.3]{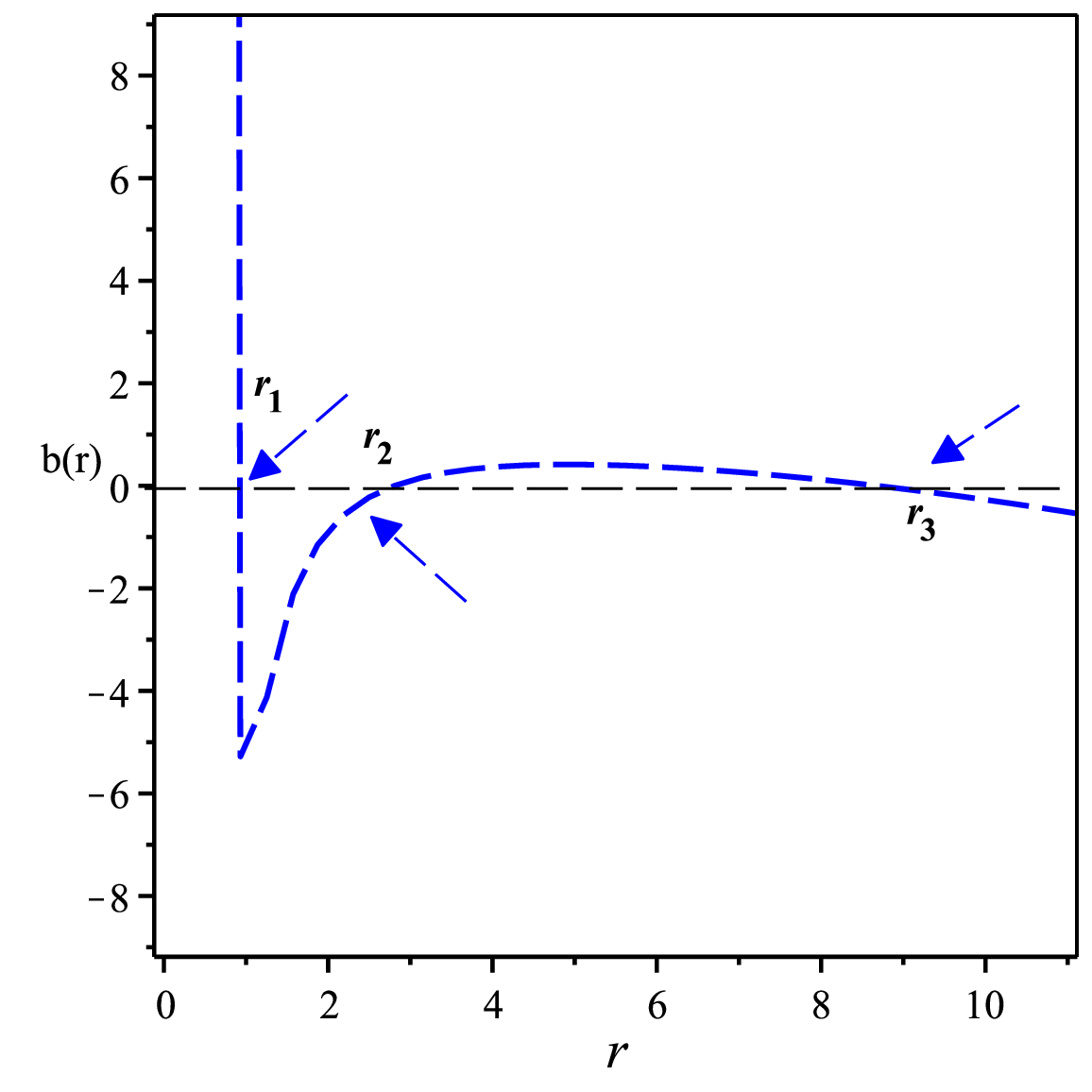}}
\subfigure[~Horizons of the solution (\ref{met11})]{\label{fig:met1}\includegraphics[scale=0.3]{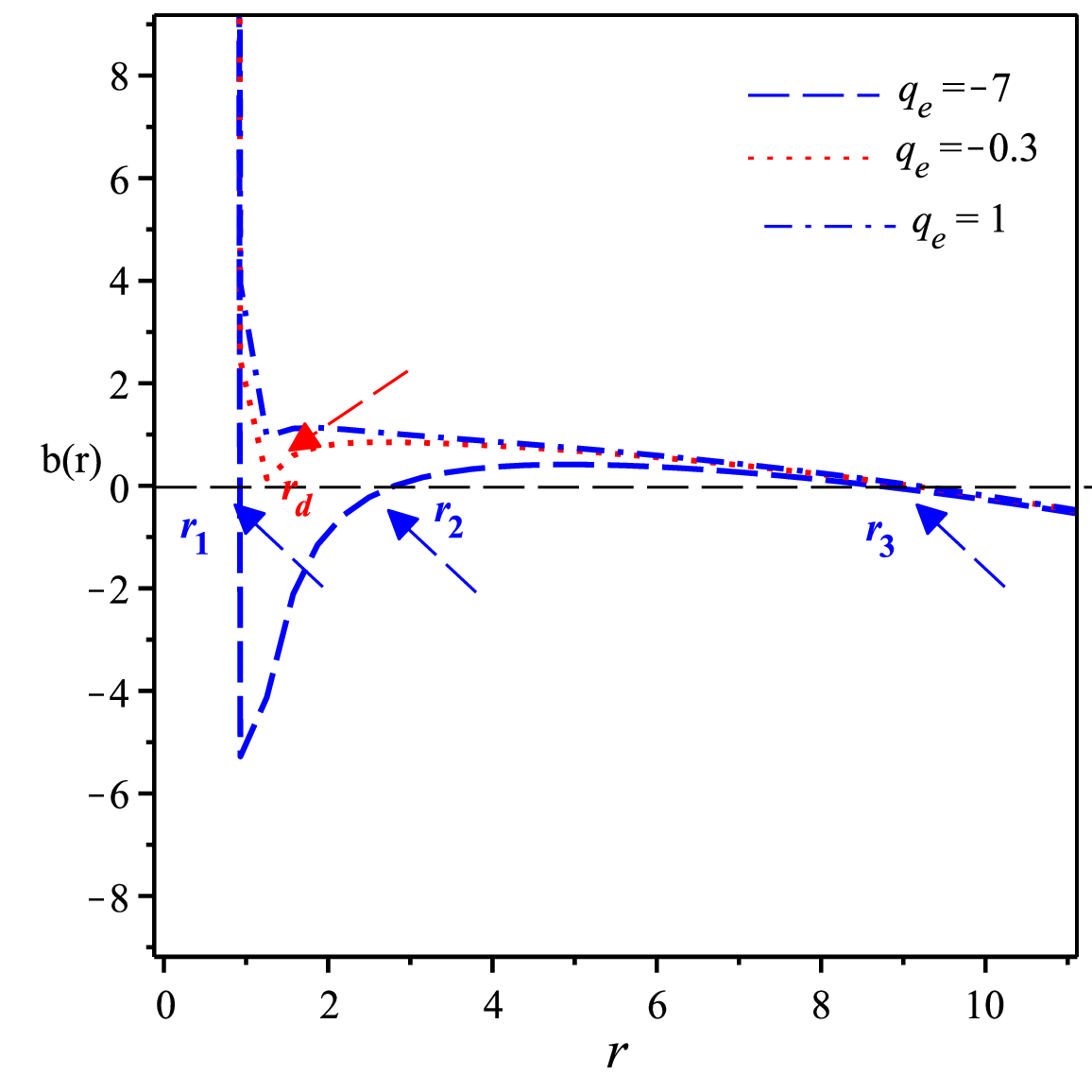}}
\subfigure[~The Hawking temperature of the BH (\ref{met11})]{\label{fig:temp}\includegraphics[scale=0.3]{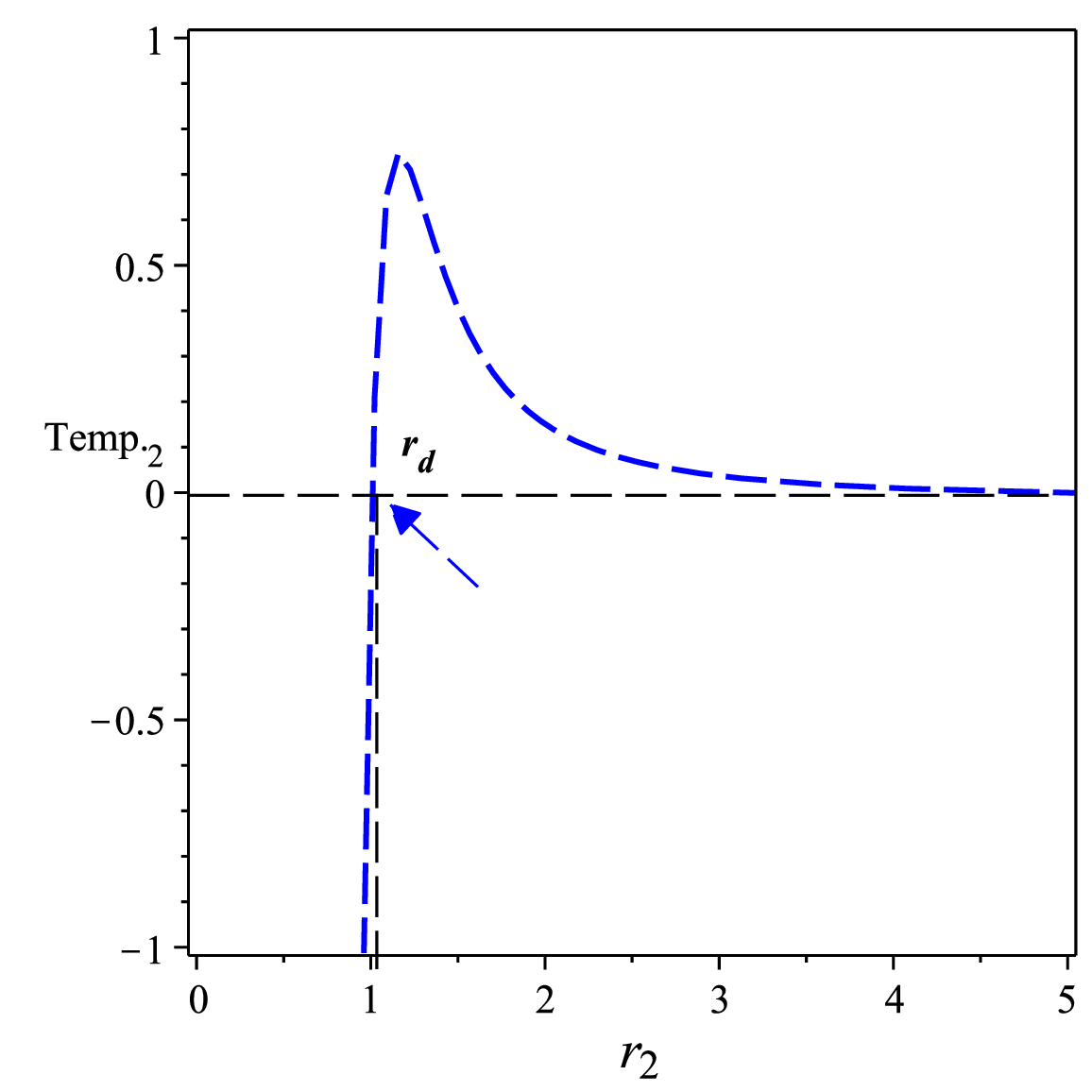}}
\subfigure[~The heat capacity of Eq.~(\ref{met11}) ]{\label{fig:ent11}\includegraphics[scale=0.3]{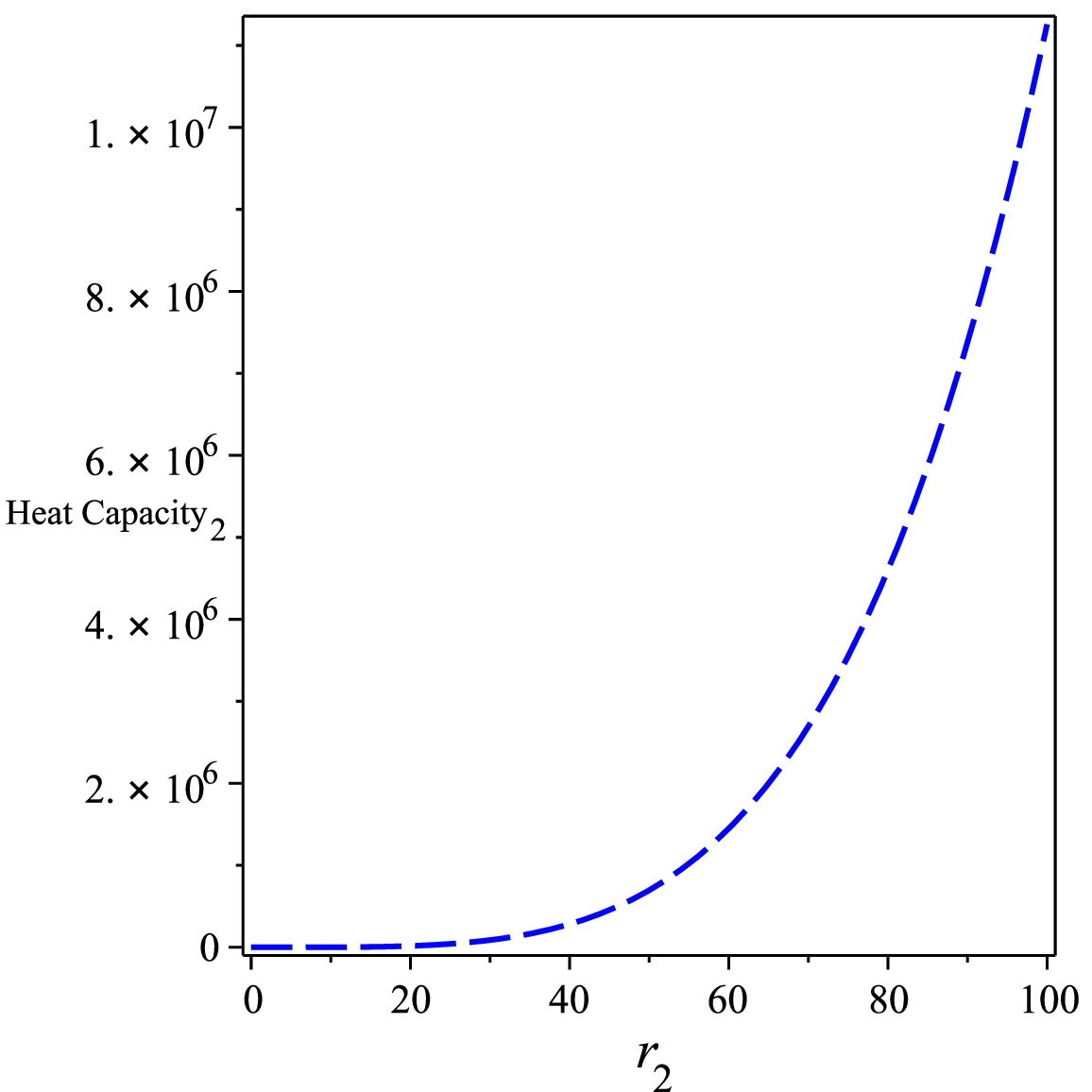}}
\subfigure[~The Gibbs energy  of Eq.~(\ref{met11})]{\label{fig:gib11}\includegraphics[scale=0.3]{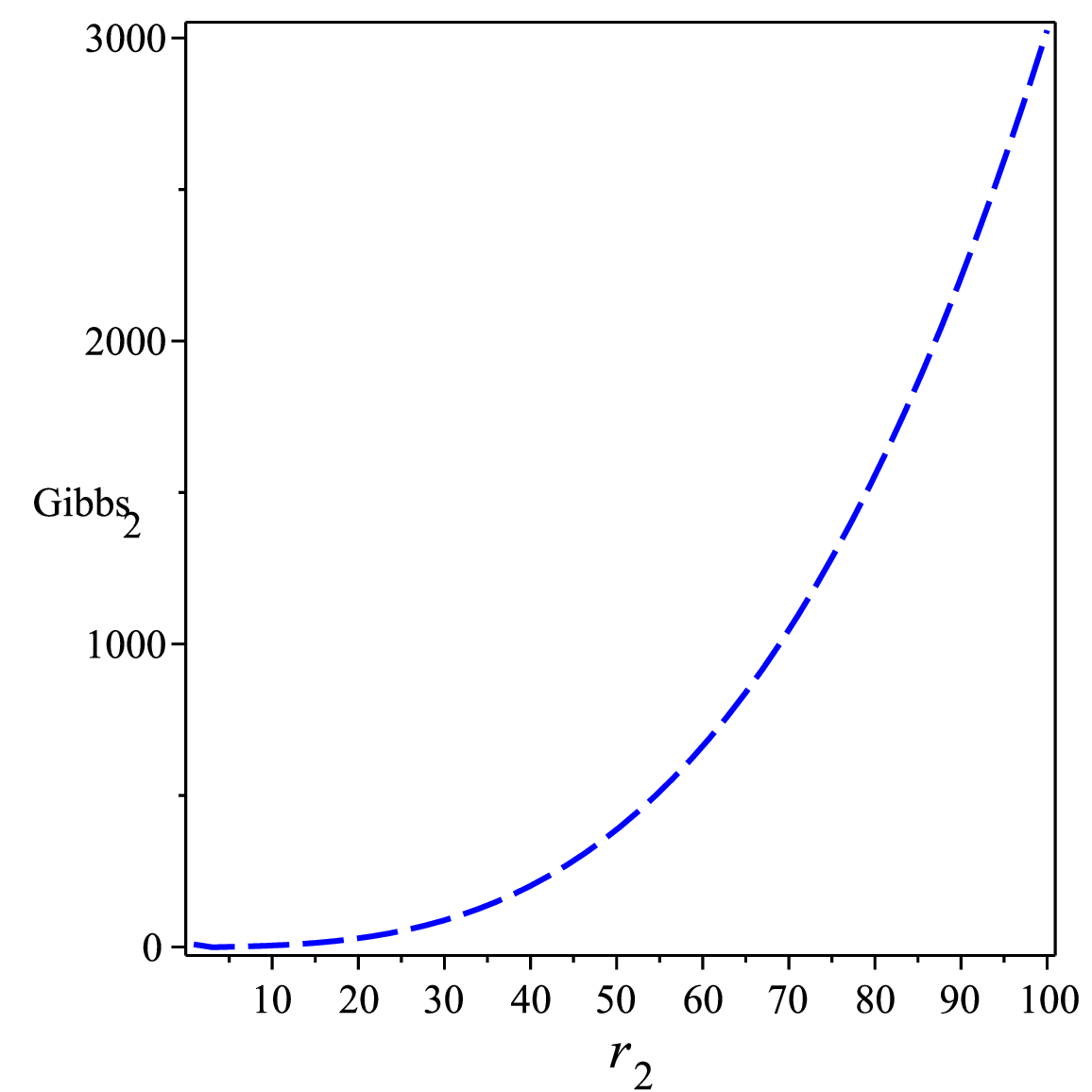}}
\caption[figtopcap]{\small{{The graph displays the thermodynamic properties of the black hole solution given by Equation (\ref{met11}). Fig.~\subref{fig:met}  represents the metric of Eq.~(\ref{met11}). On the other hand, Fig.~\subref{fig:met1}  illustrates the horizons of the metric $b$ described by Eq.~(\ref{met11}).
Fig.~\subref{fig:temp} the  temperature;
Fig.~\subref{fig:ent11} is the heat; Fig.~\subref{fig:gib11}  the Gibbs  energy.  In this study, we assign the numerical values of the parameters that define the model. Specifically, we consider $M=0.01$, $\Lambda=-0.012$, $q_e=-7$, $q_m=-7$, and $c_3=5$..}}}
\label{Fig:11}
\end{figure}

To determine the horizon radii of the black hole, as expressed in Eq.~(\ref{met11}), we solve the equation $b(r)=0$. This process yields eleven solutions. While obtaining the analytical forms of these solutions is challenging, we can numerically plot the asymptotic forms of the real solutions. Subsequently, we analyze these solutions based on the indication of the cosmological constant.\\
\\
\underline{\bf $\Lambda<0$}:\\
\\
According to the information presented in {Figure~\ref{Fig:11}~\subref{fig:met}}, Eq. (\ref{met11})   provides the BH solution. This solution exhibits three instances where the function $b(r)$ equals zero, indicating the presence of horizons. Typically, when dealing with charged (A)dS (Anti-de Sitter) BHs, one would expect to calculate two horizons. However, in this particular study, the model incorporates a third horizon due to the influence of the dimensional parameter $c_3$.

\begin{figure}
\centering
\subfigure[~The Horizons of (\ref{met11})]{\label{fig:met1p}\includegraphics[scale=0.3]{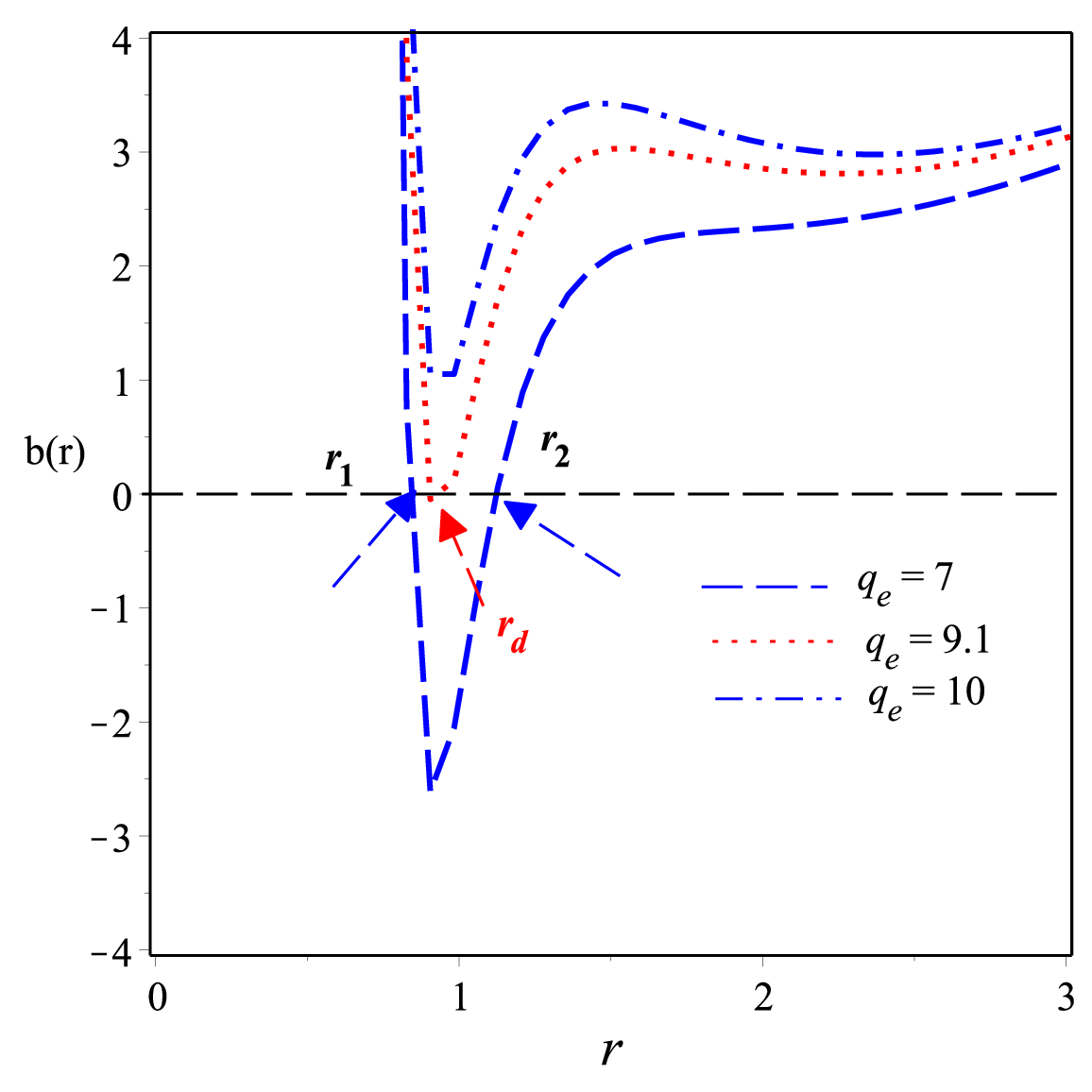}}
\subfigure[~The  temperature of the BH (\ref{met11})]{\label{fig:tempp}\includegraphics[scale=0.3]{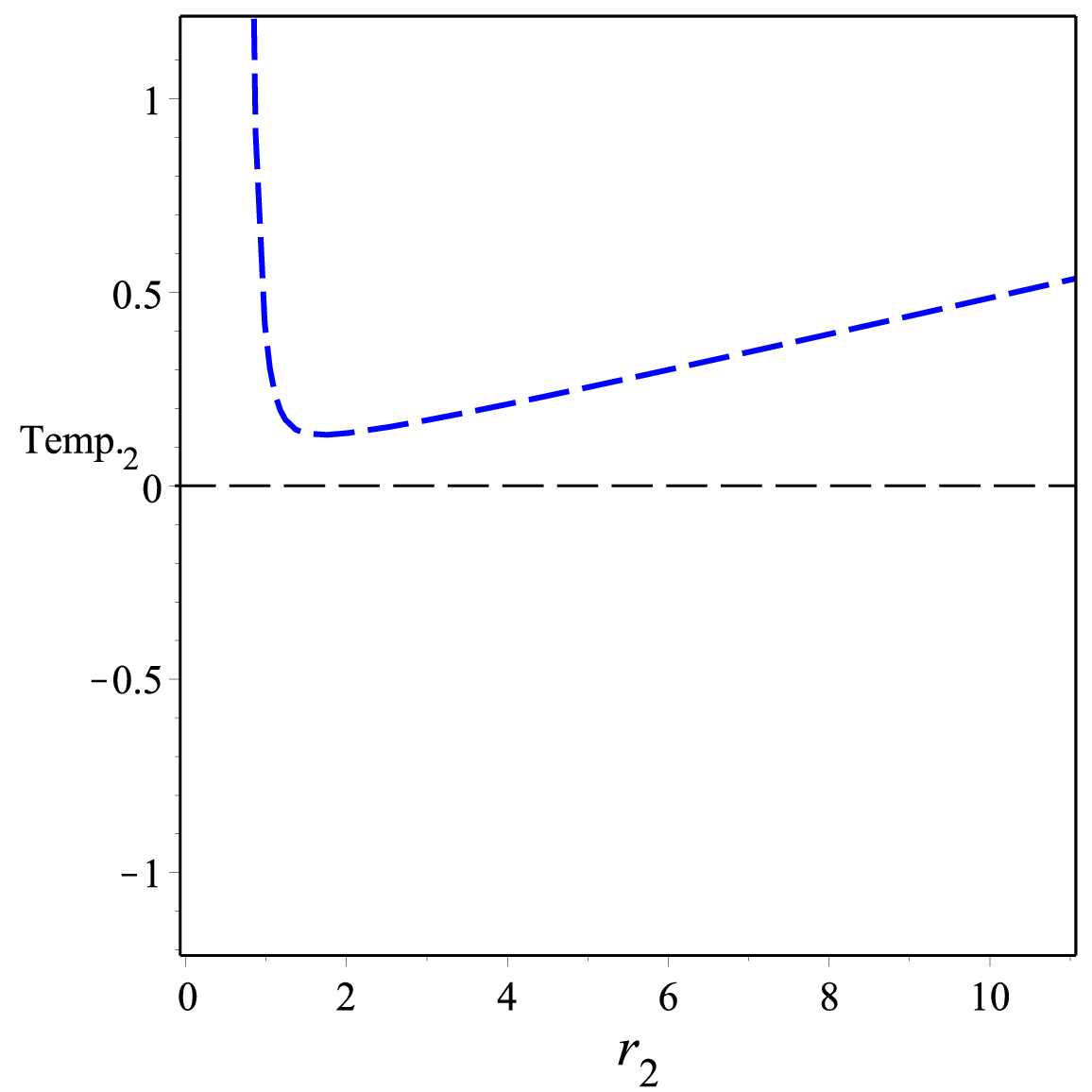}}
\subfigure[~The heat capacity of BH (\ref{met11}) ]{\label{fig:ent11p}\includegraphics[scale=0.3]{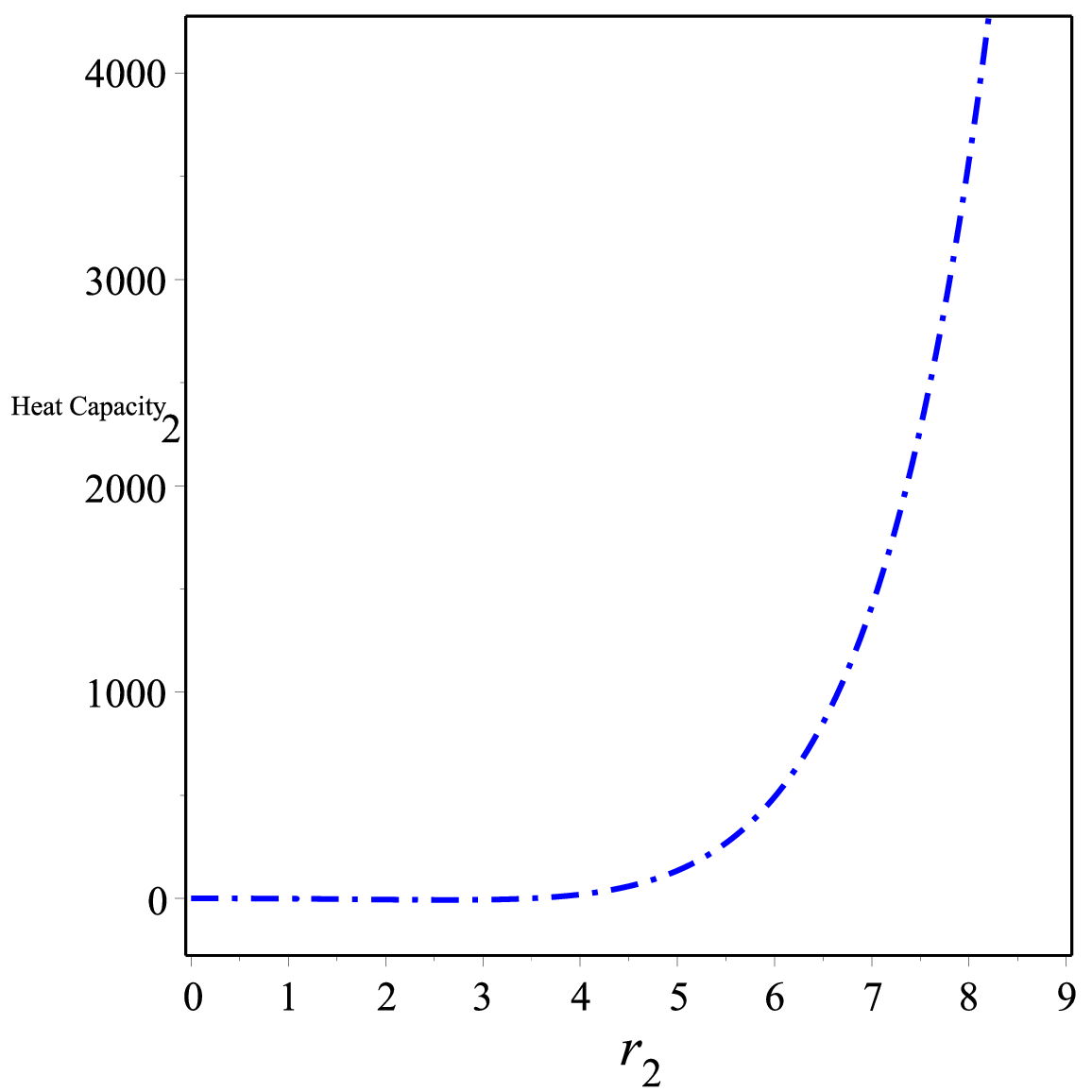}}
\subfigure[~The Gibbs  energy of Eq.~(\ref{met11})]{\label{fig:gib11p}\includegraphics[scale=0.3]{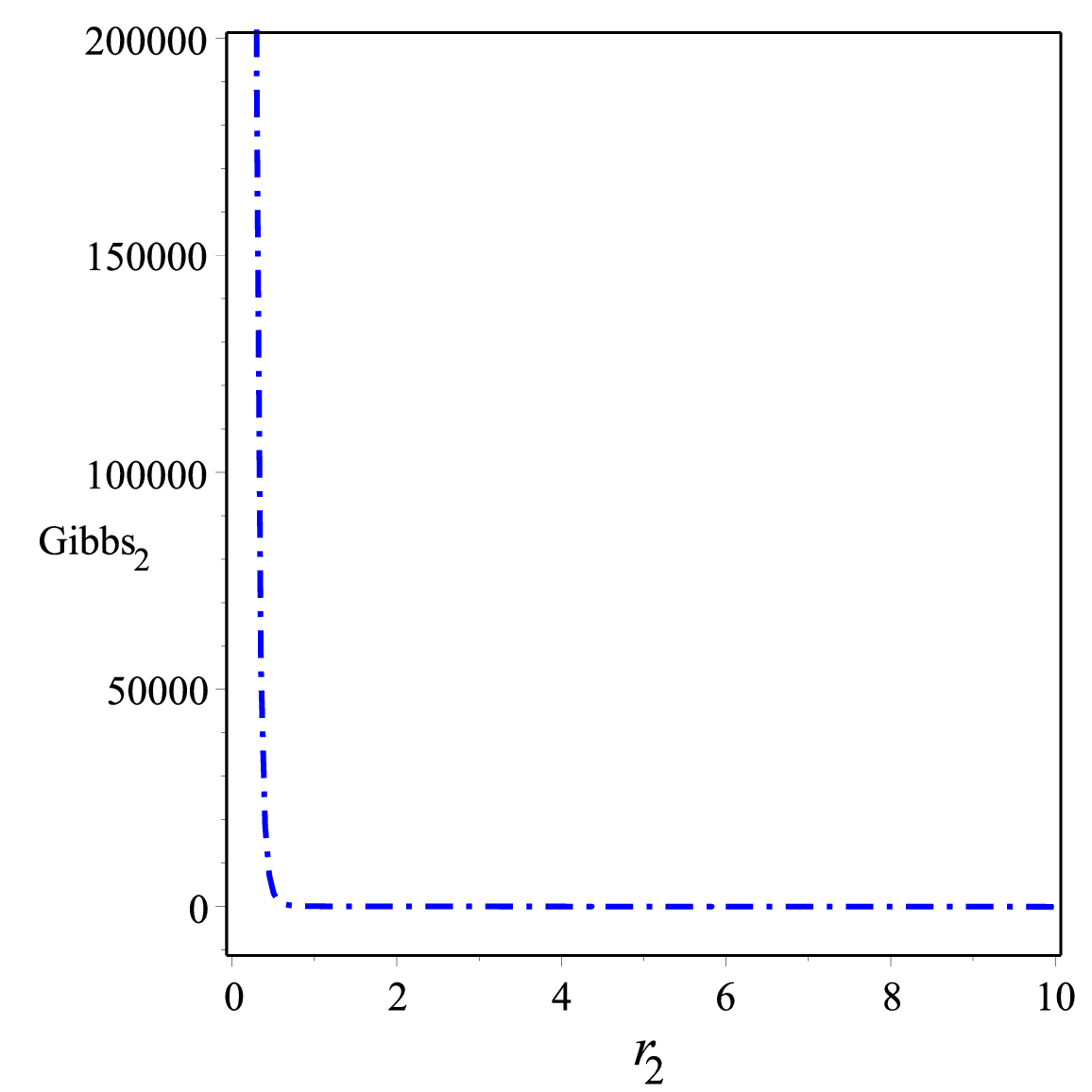}}
\caption[figtopcap]{\small{{The graph illustrates the thermodynamics of the black hole solution as described by Eq.~(\ref{met11}). Fig.~\subref{fig:met}  represents the metric of (\ref{met11}); 
In Fig.~\subref{fig:tempp}, we present the temperature. Fig.~\subref{fig:ent11p}  showcases of heat capacity. Lastly,  Figure~\subref{fig:gib11p} illustrates the Gibbs free energy. For this analysis, we utilize specific numerical values for the parameters characterizing the model, namely $M=1$, $\Lambda=0.2$, $q_e=100$, $q_m=100$, and $c_3=5$.}}}
\label{Fig:2}
\end{figure}

In the upcoming analysis, we will investigate the thermodynamic characteristics of the BH described by Eq.~(\ref{met11}).  To facilitate our investigation, we provide the fundamental definitions of thermodynamic quantities. Referring to Figure~\ref{Fig:2}~\subref{fig:met1}, we identify the two horizons, namely the inner event horizon located at $r=r_1$ and the outer horizon situated at $r=r_2$.
In the scenario where the horizons coincide, specifically when the parameters $q_e$ and $q_m$ both equal -0.3 (or more precisely $q_e=q_m=-0.3\kappa$ in units where $\kappa=1$), we observe the existence of a single horizon referred to as the degenerate horizon denoted as $r_d$. However, when the condition $q_e=q_m>-0.3$ is satisfied, a naked singularity emerges.
The Hawking temperature, which is the temperature of the black hole, is defined in the following manner:
 \cite{Sheykhi:2012zz,Sheykhi:2010zz,Hendi:2010gq,Sheykhi:2009pf,Wang:2018xhw,Zakria:2018gsf},
\begin{align}
\label{temp}
T= \frac{b'\left(r_2\right)}{4\pi}\,.
\end{align}
Moreover, the entropy of the outer horizon according to Hawking is defined as:
\begin{align}
\label{ent}
S\left(r_2\right) =\frac{1}{4} A \left(r_2 \right)\,.
\end{align}
Here $A\left(r_2\right)$ the outer horizon's area.

The  stability of any  black hole  is related to the indication of its heat capacity $H$. In order to analyze the thermal stability of the BH, we establish the heat capacities as a parameter \cite{Nouicer:2007pu,Chamblin:1999tk}:
\begin{equation}\label{m55}
H_2=\frac{\partial M}{\partial r_2} \left(\frac{\partial T}{\partial r_2}\right)^{-1}\,.
\end{equation}
Therefore, if $(H_{2} > 0, H_2 < 0)$, then the black hole is thermodynamically stable or unstable.
Furthermore, the Gibbs free energy is defined in the following manner \cite{Zheng:2018fyn, Kim:2012cma},
\begin{align}
\label{enr1}
\mathbb{G}\left(r_2\right)=M\left(r_2\right)-T\left(r_2\right)S\left(r_2\right)\,,
\end{align}
where $M\left(r_2\right)$ is given from the metric potential $b$ as:
\begin{align}\label{MM}
M\left(r_2\right)=\frac{\Lambda {r_2}^{11}+{r_2}^{9}+({ q_e}{}^2+{ q_m}^2){r_2}^{7}+c_3{r_2}^{3}+c_3{}^{3/2}}{2{r_2}^{8}}\,.
\end{align}

Using Eq.~(\ref{temp}), we calculate the Hawking temperature and get:
\begin{align}
\label{T1}
T\left(r_2\right)=\frac{3\Lambda  {r_2}^{11}+{r_2}^9-\left( {{ q_e}}^2+{{ q_m}}^2 \right)r_2{}^7-5c_3{r_2}^{3}-8c_3{}^{3/2}}{4\pi {r_2}^{10}}\,.
\end{align}
Figure~\ref{Fig:2}~\subref{fig:temp} displays the trend of the temperature calculated by Eq.~(\ref{T1}), revealing that when $r_2>r_d$, the value $T_2\equiv T\left(r_2\right)$ remains positive, { where $r_d$ is the point where the two horizons $r_1$ and $r_2$ are coincides as shown in Fig.~\ref{Fig:11}~\subref{fig:met1}  by the red curve.}
Figure \ref{Fig:1}~\subref{fig:temp} indicates that temperature $T$ vanishes when $r_2 = r_d$.
The heat capacity of the BH { given by Eq.~ (\ref{met11})} takes the following form:
\begin{align}
\label{en}
H\left( r_2 \right)=-\frac{3\Lambda  {r_2}^{11}+{r_2}^9- \left( {{ q_e}}^2+{{ q_m}}^2 \right)r_2{}^7-5c_3{r_2}^{3}-8c_3{}^{3/2}}{{r_2}^9-2 \left( {{ q_e}}^2+{{ q_m}}^2 \right)r_2{}^7-20c_3{r_2}^{3}-44c_3{}^{3/2}}\, .
\end{align}
The heat capacity's characteristics, described by Eq.~(\ref{en}), are illustrated in Figure~\ref{Fig:2}~\subref{fig:ent11}.  The plot reveals a positive value, indicating that the model, as given by Equation (\ref{met}), possesses thermodynamic stability.
Using the above thermodynamical quantities, we  get the value of Gibb's function as:
\begin{align}
\label{enr2}
\mathbb{G}\left(r_2\right)=\frac{-\Lambda  {r_2}^{11}+{r_2}^9+3\left( {{ q_e}}^2+{{ q_m}}^2 \right)r_2{}^7+7c_3{r_2}^{3}+10c_3{}^{3/2}}{4r_2{}^8}\,.
\end{align}
The Gibbs function's behavior, as represented by Eq. (\ref{enr2}), is shown in Figure~\ref{Fig:1}~\subref{fig:gib11}, demonstrating a positive value.

\underline{\bf $\Lambda>0$}:\\
\\
Using the above data Fig.~\ref{Fig:2} illustrates the dynamics of the horizons of BH (\ref{met11}), temperature, heat capacity, and Gibbs function. The scenario in which $\Lambda>0$ also leads to the development of a physical.

{ The specific gravity theory ,in the context of Gauss-Bonnet gravity coupled with a scalar field, explores the consequences of the additional terms (Gauss-Bonnet term and scalar field, potential, Lagrangian multipliers)  on various physical phenomena, including thermodynamics. The scalar field can have nontrivial interactions with curvature, leading to modifications of the first law of thermodynamics as we shall show below}.

Now we show if the 1st law of thermodynamics is satisfied for the BH { given by Eq.~(\ref{met11})} or not. To accomplish this, we take some steps to determine whether or not the first law is satisfied. Eq.~ (\ref{MM}) can be used to calculate the electric and magnetic potentials $q_e$ and $q_m$ as follows:
\begin{align}
\Phi=\left(\frac{\partial M}{\partial q_e}+\frac{\partial M}{\partial q_m}\right)_{S,P,c_3}=\frac{q_e+q_m}{r_2}\,.
\end{align}
In the extended phase space, we consider the negative of the cosmological constant as a
positive pressure term:
\begin{align}
\Phi=-\frac{3\Lambda}{4\pi}\,.
\end{align}
Such a term is interpreted as a pressure because its conjugate variable is the volume
of the sphere $S^2$ bounded by the horizon radius
\begin{align}
V=\left(\frac{\partial M}{P}\right)_{S,q_e,q_m}=\frac{4\pi{r_2}^3}{3}\,.
\end{align}
From these definitions, we can formulate the first law of thermodynamics in the extended phase
space as:
\begin{align}\label{Con}
dM=TdS+\Phi dQ+VdP+\cal{C}dc\,,
\end{align}
{ where ${\cal C}$ is the physical quantity conjugate to the parameter $c_3$, which is defined as
\begin{align}
{\cal C}=\left(\frac{\partial M}{c_3}\right)_{S,q_e,q_m,P}=\frac{2{r_2}^3+3c_3{}^{1/2}}{2{r_2}^8}\,.
\end{align}}
This way, we verify the role of the BH mass in this thermodynamic system, it
should be treated as the enthalpy of the system when we introduce the pressure the term described by the cosmological constant.

The Smarr relation  can be derived by integrating all the physical quantities
in Eq. (\ref{Con}) as:
\begin{align}
M=2TS+\Phi\,Q-VdP+3\cal{C}c_3\,.
\end{align}

\section{The stability of the aforementioned solutions using geodesic deviation}\label{S4}

Geodesic equations are given by \cite{Misner:1973prb}

\begin{equation}
\label{ge3}
\frac{d^2 x^\alpha}{d\eta^2}
+ \left\{ \begin{array}{c} \alpha \\ \beta \rho \end{array} \right\}
\frac{d x^\beta}{d\eta} \frac{d x^\rho}{d\eta}=0\, .
\end{equation}
Here the parameter $\eta$ represents the affine connection, the geodesic deviation equations can be expressed in the following manner. \cite{dInverno:1992gxs,Nashed:2003ee},
\begin{equation}
\label{ged333}
\frac{d^2 \varepsilon^\sigma}{d\eta^2}
+ 2\left\{ \begin{array}{c} \sigma \\ \mu \nu \end{array} \right\}
\frac{d x^\mu}{d\eta} \frac{d \varepsilon^\nu}{d\eta}
+ \left\{ \begin{array}{c} \sigma \\ \mu \nu \end{array} \right\}_{,\, \rho}
\frac{d x^\mu}{d\eta} \frac{d x^\nu}{d\eta}\varepsilon^\rho=0\, ,
\end{equation}
where $\varepsilon^\rho$ is the four-vector deviation. Plugging Eqs.~(\ref{ge3}) and (\ref{ged333})
into Eq.~(\ref{met11}), we get
\begin{equation}
\label{ges}
\frac{d^2 t}{d\eta^2}=0\, , \quad \frac{1}{2} b'(r) \left(
\frac{d t}{d\eta}\right)^2 - r\left( \frac{d \phi}{d\eta}\right)^2=0\, , \quad
\frac{d^2 \theta}{d\eta^2}=0 \, ,\quad \frac{d^2 \phi}{d\eta^2}=0\, .
\end{equation}
As for the geodesic deviation, the line-element can be described by the following expression
\begin{align}
\label{ged11}
& \frac{d^2 \varepsilon^1}{d\eta^2} +b(r)b'(r) \frac{dt}{d\eta}
\frac{d \varepsilon^0}{d\eta} -2r b(r) \frac{d \phi}{d\eta}\frac{d \varepsilon^3}{d\eta}
+\left[ \frac{1}{2} \left( b'^2(r)+b(r) b''(r)
\right)\left( \frac{dt}{d\eta} \right)^2-\left(b(r)+rb'(r)
\right) \left( \frac{d\phi}{d\eta}\right)^2 \right]\varepsilon^1=0\, ,
\nonumber\\
& \frac{d^2 \varepsilon^0}{d\eta^2} + \frac{b'(r)}{b(r)} \frac{dt}{d\eta}
\frac{d {\varepsilon^1}}{d\eta}=0\, ,\quad \frac{d^2 \varepsilon^2}{d\eta^2}
+\left( \frac{d\phi}{d\eta}\right)^2 \varepsilon^2=0\, , \quad
\frac{d^2 \varepsilon^3}{d\eta^2} + \frac{2}{r} \frac{d\phi}{d\eta}
\frac{d \varepsilon^1}{d\eta}=0\, ,
\end{align}
where $b(r)$ is the solution given  Eqs~(\ref{sol4}). The prime symbol, denoting the derivative with respect to the radial coordinate $r$, is utilized. When considering a circular orbit, the following outcomes are obtained.
\begin{equation}
\label{so}
\theta= \frac{\pi}{2}\, , \quad
\frac{d\theta}{d\eta}=0\, , \quad \frac{d r}{d\eta}=0\, .
\end{equation}
When Eq.(\ref{so}) is employed in Eq.(\ref{ges}), it results in the following outcome
\begin{equation}
\left( \frac{d\phi}{d\eta}\right)^2= \frac{b'(r)}{r \left[ 2b(r)-rb'(r) \right]}\, , \quad
\left( \frac{dt}{d\eta}\right)^2= \frac{2}{2b(r)-rb'(r)}\, .
\end{equation}

Eq.~(\ref{ged11}) can be rewritten in the following form
\begin{align}
\label{ged2222}
& \frac{d^2 \varepsilon^1}{d\phi^2}
+b(r)b'(r) \frac{dt}{d\phi} \frac{d \varepsilon^0}{d\phi}
 -2r b(r) \frac{d \varepsilon^3}{d\phi} +\left[ \frac{1}{2} \left[b'(r)^2+b(r) b''(r)
\right]\left( \frac{dt}{d\phi}\right)^2-\left[b(r)+rb'(r) \right] \right]\zeta^1=0\, , \nonumber\\
& \frac{d^2 \varepsilon^2}{d\phi^2}+\varepsilon^2=0\, , \quad
\frac{d^2 \varepsilon^0}{d\phi^2} + \frac{b'(r)}{b(r)}
\frac{dt}{d\phi} \frac{d \varepsilon^1}{d\phi}=0\, , \quad \frac{d^2 \varepsilon^3}{d\phi^2}
+ \frac{2}{r} \frac{d \varepsilon^1}{d\phi}=0\, .
\end{align}
 By considering the second equation in Eq.(\ref{ged2222}), it can be demonstrated that we obtain a simple harmonic motion, which corresponds to the stability condition of the plane $\theta=\pi/2$. The remaining equations, as provided by Eq.(\ref{ged2222}), have the following solutions.
\begin{equation}
\label{ged33}
\varepsilon^0 = \zeta_1 \e^{i \sigma \phi}\, , \quad
\varepsilon^1= \zeta_2\e^{i \sigma \phi}\, , \quad \mbox{and} \quad
\varepsilon^3 = \zeta_3 \e^{i \sigma \phi}\, .
\end{equation}
Here, $\zeta_1$, $\zeta_2$, and $\zeta_3$ represent constants, while $\omega$ is an unknown function. By substituting the values of $\varepsilon^1$ and $\varepsilon^3$ from Eq.(\ref{ged33}) into the fourth equation of Eq.(\ref{ged2222}), we obtain the following result.
\begin{equation}
\label{c1}
\zeta_2=-\frac{\zeta_3\sigma\,r}{2}\,.
\end{equation}
After substituting the values of $\varepsilon^0$ and $\varepsilon^1$ from Eq.(\ref{ged33}) into the third equation of Eq.(\ref{ged2222}), and subsequently utilizing Eq.~(\ref{c1}), we obtain the following outcome.
\begin{equation}
\label{c2}
\zeta_1=\frac{\zeta_3\sqrt{2b'\,r^3}}{2b}\,.
\end{equation}
Substituting {  Eq.~(\ref{ged33})}, after using Eqs.~(\ref{c1}) and (\ref{c2}), into the first equation of Eq.~(\ref{ged2222})
we obtain the stability condition as
\begin{equation}
\label{con111}
\frac{3b b'-\sigma^2 b'-2r{b'}^{2}+rb b'' }{b'}>0\Rightarrow \sigma^2<\frac{3bb'-2r{b'}^{2}+rbb'' }{b'}\, .
\end{equation}
\begin{figure}
\centering
\subfigure[~The stability condition for $\Lambda<0$]{\label{fig:stab}\includegraphics[scale=0.3]{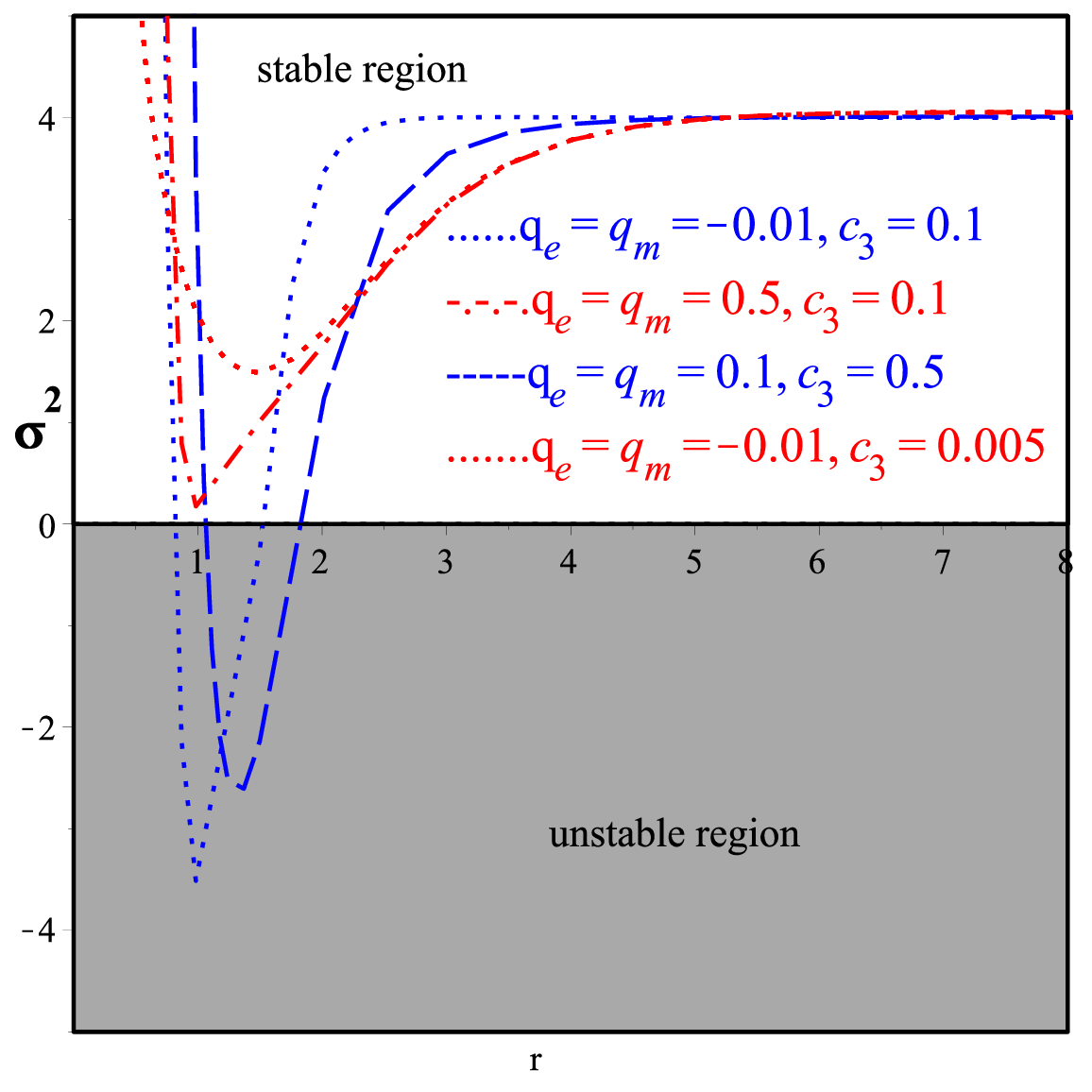}}
\subfigure[~The stability condition for $\Lambda>0$]{\label{fig:stab1}\includegraphics[scale=0.3]{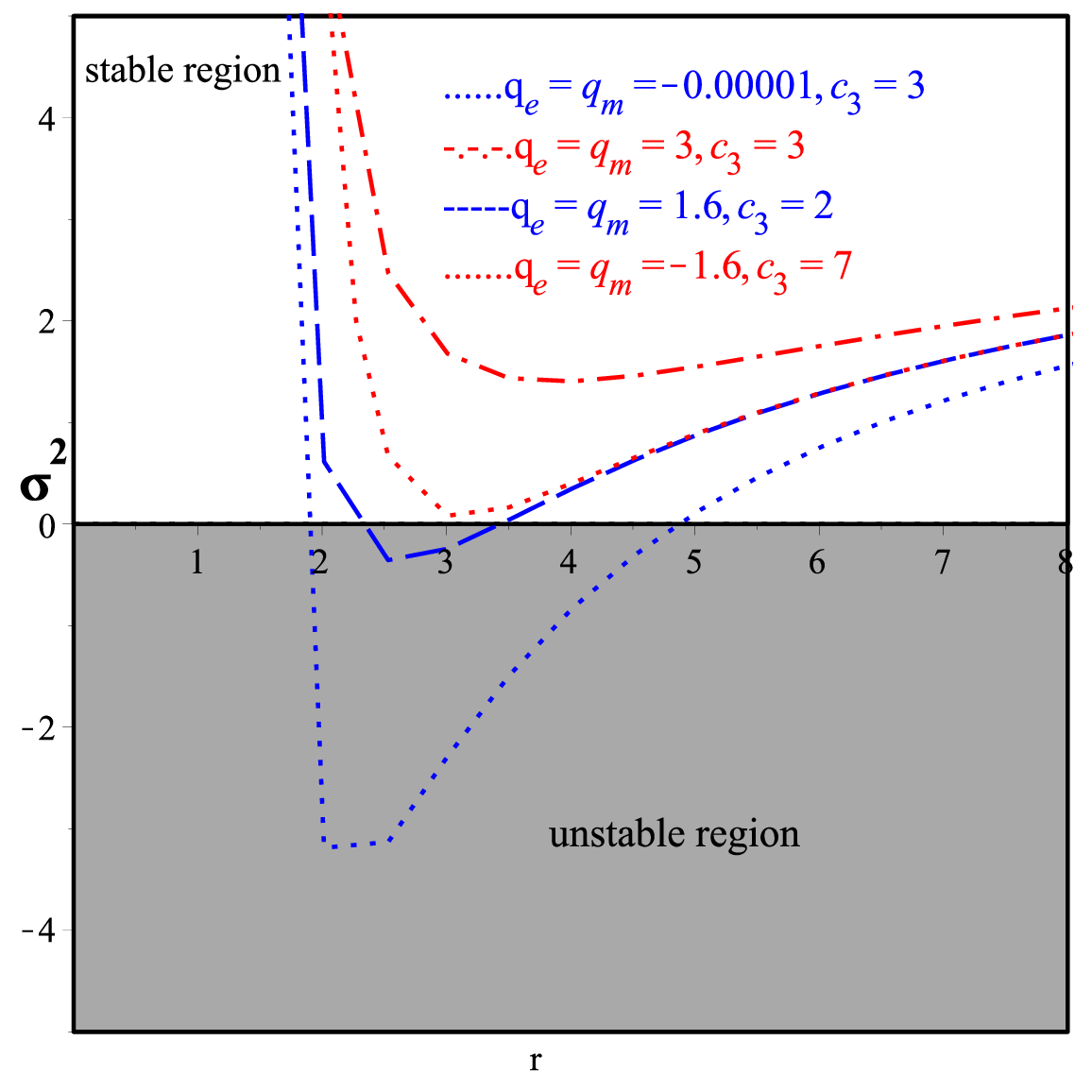}}
\caption[figtopcap]{\small{{Plot of the stability condition given by Eq. (\ref{con111}).  Figure~\subref{fig:stab} is the stability condition of the BH { given by Eq.~ (\ref{con111})} when $\Lambda=-0.012$ and  $M=0.00011$; Figure~\subref{fig:stab1} is the stability condition of the BH { given by Eq.~ (\ref{con111})} when $\Lambda=3$ and $M=1.3$. }}}
\label{Fig:3}
\end{figure}
We depict Eq.~(\ref{con111}) in Fig.~\ref{Fig:3} using specific values of the model. The figures exhibit the unshaded and shaded zones where
the BHs are stable and not stable.
\section{Conclusion and discussion}\label{S5}
Charged BHs are important because they have unique properties that make them different from neutral BHs. In this study we have derived a new non-trivial charged BH solution employing the combination of the Gauss-Bonnet gravitational theory with a scalar field \cite{Nojiri:2018ouv, Nojiri:2021mxf}. 
 For this aim, we have embraced the effect of electric and magnetic charges in the equation of motions, adding the influence of Maxwell's to Gauss-Bonnet gravitational theory with a scalar field electromagnetic field. We are interested in the linear form of the Maxwell field equation, and the non-linear case will be studied elsewhere.

We used a spherically symmetric spacetime with a single unknown function where $g_{tt}=g_{rr}$. By imposing the electric and magnetic charges as well as the cosmological constant, $\Lambda$, we get a system of non-linear differential equations. We were able to derive the form of magnetic fields by using this    system  of off-diagonal components, and we were able to derive the form of an electric field by using the form of the Maxwell field. Using these data in the diagonal form of the field equations, we succeeded in deriving the metric potential that characterized the spacetime under consideration as well as the parameters that characterized Gauss-Bonnet Â gravitational theory coupled to a scalar field. The main characterization of this solution is the fact that the metric is affected by the electric and magnetic fields as expected, in addition to the fact that it has extra terms of order, $O(\frac{1}{r^6})$ and $O(\frac{1}{r^9})$.

We examined this solution by studying its horizons and showed that if $\Lambda<0$, we have three horizons. { Usually, for a neutral BH there are two horizons when one has a solution with a cosmological constant; in our solution, we created three horizons for a charged BH solution.} The third horizon is primarily caused due to the higher order appeared in the metric potential, $O(\frac{1}{r^6})$ and $O(\frac{1}{r^9})$, which are related to the scalar field, If $\Lambda>0$ we showed that there are only two horizons which are Cauchy and outer horizons.

We also examined our BH using thermodynamical quantities. We calculated the Hawking temperature and showed that we have a positive one as far as $r>r_d$ where $r_d$ is the degenerate horizon. Also, we calculated the heat capacity and Gibbs function of the BH under consideration and showed that we have a stable model regardless of the sign of the cosmological constant. Also, we derived the modified form of the first law and the Smarr relation, which insures the validity of the first law of thermodynamics derived in this study.

As a final physical test of the BH given by Eq.~(\ref{met11})  we studied the geodesic deviation and derived the condition of stability. This condition is analyzed graphically to show the stable and unstable regions when $\Lambda>0$ and $\Lambda<0$. As far as this is the first time to obtain an electrified analytical solution within the context of the Gauss-Bonnet gravitational theory linked to the scalar field while utilizing identical metric potential. The case with unequal metric potential, $g_{tt}\neq g_{rr}$ will be studied elsewhere. Moreover, the geodesic of the BH  given by Eq.~(\ref{met11}) should be studied to check if the extra terms, $O(\frac{1}{r^6})$ and $O(\frac{1}{r^9})$, affect the paths and are consistent with the recent gravitational waves data \cite{LIGOScientific:2017zic,LIGOScientific:2017bnn}. This will be studied in our future project.

{
\centerline{\bf Appendix}
The explicate form of $\omega$, $\lambda$, $h(r)$ and ${\tilde V}$ have the following form:
\begin{align}
\label{sol441}
 & \omega(r)=-\frac{\mu^2}{c_3{}^2b(r)}\,,\end{align}
\begin{align} \label{sol44}
&\lambda=-16\, \bigg[ -1/2\,exp\bigg\{\int \bigg\{ -2\,{r}^{20}\Lambda-3\,c_4\,{r}^{19}\Lambda+4\,{r}^{17}c_4+ \left( 6\,{c_4}^{2}+10\,c_1 \right) {r}^{16}+21\,c_4\, {r}^{15}c_1+ \left( 16\,{c_1}^{2}+32\,\Lambda\,c_3 \right) {r}^{14}\nonumber\\
&+54\,{r}^{12}c_3+ \left( 77\,c_4\,c_3+ 77\,\Lambda\,c_3{}^{3/2} \right) {r}^{11}+96\,c_1\,{r}^{10}c_3+ 108\,{r}^{9}c_3{}^{3/2}+140\,c_4\,{r}^{8}c_3{}^{3/2}+165\,c_1\,{r} ^{7}c_3{}^{3/2}+96\,{c_3}^{2}{r}^{6}\nonumber\\
&+285\,{r}^{3}c_3{}^{5/2}+ 198\,{c_3{}^{3}}\bigg\}\bigg[ \left( 3c_4{r}^{8}+4c_1{r}^{7} +8c_3{r}^{3}+11c_3{}^{3/2} \right) r \left( \Lambda{r}^{11}+ {r}^{9}+c_4{r}^{8}+c_1{r}^{7}+c_3{r}^{3}+c_3{}^{3/2} \right) \bigg]^{-1}{dr}\bigg\} \nonumber\\
&\bigg( {r}^{20}\Lambda+3/2c_4\,{r}^{19} \Lambda-2{r}^{17}c_4-21/2c_4\,{r}^{15}c_1+ \left( -5\,c_1-3\,{c_4}^{2} \right) {r}^{16}+ \left( -16\, \Lambda\,c_3-8\,{c_1}^{2} \right) {r}^{14}-27\,{r}^{12}c_3- \bigg( {\frac {77}{2}}\,\Lambda\,c_3{}^{3/2}\nonumber\\
&+{\frac {77}{2}}\,c_4\,c_3 \bigg) {r}^{11}-48\,c_1\,{r}^{10}c_3-54\,{r}^ {9}c_3{}^{3/2}-70\,c_4\,{r}^{8}c_3{}^{3/2}-{\frac {165}{2}}c_1 {r}^{7}c_3{}^{3/2}-48{c_3}^{2}{r}^{6}-{\frac {285}{2}} {r}^{3}c_3{}^{5/2}-99\,{c_3{}^{3}} \bigg) \nonumber\\
&\int \! \left(  \left( c_1+{c_1}^{2}+{c_2}^{2} \right) {r}^{7}+10\,c_3\,{r }^{3}+22\,c_3{}^{3/2} \right) {r}^{9}exp\bigg\{-\int \bigg\{-2\,{r}^{ 20}\Lambda-3\,c_4\,{r}^{19}\Lambda+4\,{r}^{17}c_4+ \left( 6 \,{c_4}^{2}+10\,c_1 \right) {r}^{16}+21\,c_4\,{r}^{15}c_1\nonumber\\
&+ \left( 16\,{c_1}^{2}+32\,\Lambda\,c_3 \right) {r}^{ 14}+54\,{r}^{12}c_3+ \left( 77c_4c_3+77\Lambda\,c_3{}^{3/2} \right) {r}^{11}+96c_1{r}^{10}c_3+108{r}^{9}c_3{}^{3/2}+140\,c_4\,{r}^{8}c_3{}^{3/2}+165\,c_1\,{r}^{7}c_3{}^{3/2}\nonumber\\
&+ 96{c_3}^{2}{r}^{6}+285\,c_3{r}^{3}c_3{}^{3/2}+198{c_3{}^{3}}\bigg\}\bigg[ \left( 3c_4{r}^{8}+4c_1{r}^{7}+8c_3 {r}^{3}+11c_3{}^{3/2} \right) r \bigg( \Lambda{r}^{11}+{r}^{9}+c_4{r}^{8}+c_1{r}^{7}+c_3{r}^{3}+c_3{}^{3/2} \bigg) \bigg]^{-1}{dr}\nonumber\\
&\left( \Lambda\,{r}^{11}+{r}^{9}+c_4\,{r}^{8}+c_1\, {r}^{7}+c_3\,{r}^{3}+c_3{}^{3/2} \right) ^{-1} \left( 3\,c_4\,{ r}^{8}+4\,c_1\,{r}^{7}+8\,c_3\,{r}^{3}+11\,c_3{}^{3/2} \right) ^{-1}{dr}\bigg\}+ \bigg( {r}^{20}\Lambda-2\,{r }^{17}c_4+3/2\,c_4\,{r}^{19}\Lambda \nonumber\\
&+ \left( -5\,c_1-3\,{c_4}^{2} \right) {r}^{16} -21/2\,c_4\,{r}^{15}c_1+ \left( -16\,\Lambda\,c_3-8\,{c_1}^{2} \right) {r}^{14}-27\,{r}^{12}c_3+ \bigg( -{\frac {77 }{2}}\,\Lambda\,c_3{}^{3/2}-{\frac {77}{2}}\,c_4\,c_3 \bigg) { r}^{11}-54\,{r}^{9}c_3{}^{3/2}\nonumber\\
&-48\,c_1\,{r}^{10}c_3-70c_4\,{r}^{8}c_3{}^{3/2}-{\frac {165}{2}}c_1\,{r}^{7}c_3{}^{3/2}-48\,{c_3}^{2}{r}^{6}-{\frac {285}{2}}c_3\,{r}^{3}c_3{}^{3/2}-99{c_3{}^{3}} \bigg)c_5\,exp\bigg\{\int\bigg\{-2{r}^{20} \Lambda-3c_4{r}^{19}\Lambda+4{r}^{17}c_4\nonumber\\
&+ \bigg( 6{c_4}^{2}+10\,c_1 \bigg) {r}^{16} +21\,c_4\,{r}^{15}c_1+ \left( 16\,{c_1}^{2}+32\,\Lambda\,c_3 \right) {r}^{14}+ 54\,{r}^{12}c_3+\left( 77c_4\,c_3+77\Lambda\,c_3{}^{3/2} \right) {r}^{11}+96\,c_1{r}^{10}c_3+108{r}^{9}c_3{}^{3/2} \nonumber\\
&+140c_4{r}^{8}c_3{}^{3/2}+165c_1\,{r}^{7}c_3{}^{3/2}+96 {c_3}^{2}{r}^{6}+285c_3{r}^{3}c_3{}^{3/2}+198{c_3{}^{3/2}}^{ 2}\bigg\}\bigg[ \left( 3c_4{r}^{8}+4\,c_1\,{r}^{7}+8\,c_3\,{r} ^{3}+11\,c_3{}^{3/2} \right) r\nonumber\\
& \bigg( \Lambda{r}^{11}+{r}^{9}+c_4 {r}^{8}+c_1\,{r}^{7}+c_3\,{r}^{3}+c_3{}^{3/2} \bigg) \bigg]^{-1}{dr}\bigg\} +1/4\,{r}^{10} \bigg(  \left( c_1+{c_1}^{2}+{c_2}^{2} \right) {r}^{7}+10\,c_3\,{r}^{3}+22\,c_3{}^{3/2} \bigg)  \bigg]\nonumber\\
&\left( c_3\,{r}^{3}+c_1\,{r}^{7}+\Lambda\,{r}^{11}+c_4 \,{r}^{8}+c_3{}^{3/2} \right) {\mu}^{-4}{r}^{-21} \left( 3\,c_4\,{r} ^{8}+4\,c_1\,{r}^{7}+8\,c_3\,{r}^{3}+11\,c_3{}^{3/2} \right) ^{ -1}\,,\end{align}
\begin{align}\label{sol444}
&h  =-1/2\,\int \! \bigg[ \int \bigg[  \left( c_1+{c_1}^{2}+{c_2}^{2} \right) {r}^{7}+10\,c_3\,{r}^{3 }+22\,c_3{}^{3/2} \bigg] {r}^{9}\bigg(exp\bigg\{-\int \bigg \{-2\,{r}^{20} \Lambda-3\,c_4\,{r}^{19}\Lambda+4\,{r}^{17}c_4+ \bigg( 6\,{c_4}^{2}\nonumber\\
&+10\,c_1 \bigg) {r}^{16}+21\,c_4\,{r}^{15}c_1+ \left( 16\,{c_1}^{2}+32\,\Lambda\,c_3 \right) {r}^{14}+ 54\,{r}^{12}c_3+ \left( 77\,c_4\,c_3+77\,\Lambda\,c_3{}^{3/2} \right) {r}^{11}+96\,c_1\,{r}^{10}c_3+108\,{r}^{9}c_3{}^{3/2}\nonumber\\
&+140\,c_4\,{r}^{8}c_3{}^{3/2}+165\,c_1\,{r}^{7}c_3{}^{3/2}+96\, {c_3}^{2}{r}^{6}+285\,c_3\,{r}^{3}c_3{}^{3/2}+198\,{c_3{}^{3/2}}^{ 2}\bigg\}\bigg[ \left( 3\,c_4\,{r}^{8}+4\,c_1\,{r}^{7}+8\,c_3\,{r} ^{3}+11\,c_3{}^{3/2} \right) r \nonumber\\
&+ \left( \Lambda\,{r}^{11}+{r}^{9}+c_4 \,{r}^{8}+c_1\,{r}^{7}+c_3\,{r}^{3}+c_3{}^{3/2} \right) \bigg]^{-1}{dr}\bigg\} \bigg)\left( \Lambda\,{r}^{11}+{r}^{9}+c_4\,{r}^{8}+c_1\,{r}^{7} +c_3\,{r}^{3}+c_3{}^{3/2} \right) ^{-1}-2\,c_5\nonumber\\
&  \left( 3\,c_4\,{r}^{8}+ 4\,c_1\,{r}^{7}+8\,c_3\,{r}^{3}+11\,c_3{}^{3/2} \right) ^{-1}{d r} \bigg] \bigg\{exp\bigg\{\int \bigg\{4{r}^{17}c_4\Lambda-2\,{r}^{20}- 3c_4{r}^{19}\Lambda+ \left( 6{c_4} ^{2}+10c_1 \right) {r}^{16}+21c_4{r}^{15}c_1+54 {r}^{12}c_3\nonumber\\
&+ \left( 16\,{c_1}^{2}+32\,\Lambda\,c_3 \right) {r}^{14}+ \left( 77\,c_4\,c_3+77\,\Lambda\,c_3{}^{3/2} \right) {r}^{11}+96c_1{r}^{10}c_3+108\,{r}^{9}c_3{}^{3/2}+ 140\,c_4\,{r}^{8}c_3{}^{3/2}+165\,c_1\,{r}^{7}c_3{}^{3/2}+96\,{c_3}^{2}{r}^{6}\nonumber\\
&+285\,c_3{r}^{3}c_3{}^{3/2}+198{c_3{}^{3}} \bigg\}\bigg[ \bigg( 3c_4{r}^{8}+4\,c_1{r}^{7}+8\,c_3{r}^{ 3}+11c_3{}^{3/2} \bigg) r \bigg( \Lambda{r}^{11}+{r}^{9}+c_4{ r}^{8}+c_1{r}^{7}+c_3{r}^{3}+c_3{}^{3/2} \bigg) \bigg]^{-1}{dr}\bigg\}\bigg\}{d r}+c_6
\,,\end{align}
\begin{align}\label{sol4444}
&{\tilde V}  = \bigg\{ 8\, \bigg[ -96\,c_1\,{r}^{10}c_3{}^{3/2} \,c_3-12\,c_4\,{r}^{26}c_1\,\Lambda+ \left( -3\,{c_3}^{2}-{\frac {167}{2}}\,c_4\,c_3{}^{3/2}\,c_1 \right) {r}^{ 15}+ \left( -64\Lambda c_3{}^{3/2}-16c_4c_3 \right) {r}^ {20}- \bigg( {\frac {199}{2}}\,c_4c_3{}^{3/2}\Lambda\nonumber\\
&+{\frac {79 }{4}}{c_4}^{2}c_3+12\,c_1\,c_3 \bigg) {r}^{19}+ \left(  \left( -50\,\Lambda\,c_3-5\,{c_1}^{2} \right) c_4-55\,{\Lambda}^{2}c_3{}^{3/2} \right) {r}^{22}+ \left( -2\,{c_1}^ {3}-52\,c_1\,c_3\,\Lambda \right) {r}^{21}- \bigg( {\frac { 169}{4}}\,{c_4}^{2}c_3{}^{3/2} \nonumber\\
&+30\,c_1\,c_3{}^{3/2} \bigg) {r}^{ 16}- \bigg(34\,\Lambda\,c_3{c_1}^{2}+{\frac {15}{4}}\,{c_4}^{2}c_1 \bigg) {r}^{23}- \left( 122\,c_3\,c_3{}^{3/2} \,\Lambda+41\,{c_1}^{2}c_3{}^{3/2}+31\,c_4\,{c_3}^{2} \right) {r}^{14}+ \bigg( -1/2\,{c_4}^{2}-28\,{\Lambda}^{2}c_3 \nonumber\\
&-8\,{c_1}^{2}\Lambda \bigg) {r}^{25}+ \left( -2\,c_4\,c_1-3/4\,{c_4}^{3} \right) {r}^{24}+ \left( -20\,{c_1}^{ 2}c_3-37c_4c_3{}^{3/2}-40\,\Lambda\,{c_3}^{2} \right) {r}^{17}+ \left( -{\frac {319}{4}}\,{c_3{}^{3}}\Lambda-{\frac {199} {2}}\,c_4\,c_3{}^{3/2}\,c_3 \right) {r}^{11} \nonumber\\
&+ \left( -9/2\,{c_3{}^{2}}-12{c_3}^{3} \right) {r}^{9}- \left( 8c_1 \Lambda+{\frac {15}{4}}\,\Lambda\,{c_4}^{2} \right) {r}^{27}+{r}^ {31}{\Lambda}^{2}-{\frac {99}{4}}{c_3{}^{9/2}}-6{\Lambda}^{2}{r}^ {29}c_1-40\left( c_4\,c_3+5/2\Lambda\,c_3{}^{3/2} \right) c_1{r}^{18}\nonumber\\
&-30c_1{r}^{13}{c_3}^{2}-12c_3{r}^{12}c_3{}^{3/2}-51{c_3}^{2}{r}^{6}c_3{}^{3/2}-{\frac { 255}{4}}c_3{r}^{3}{c_3{}^{3}}-{\frac {265}{4}}c_4\, {r}^{8}{c_3{}^{3}} -{\frac {255}{4}}c_1\,{r}^{7}{c_3{}^{3}}-3{\Lambda}^{2}{r}^{30}c_4-4\,\Lambda c_4{r}^{28} \bigg] \nonumber\\
&exp\bigg\{\int \bigg\{4{r}^{17}c_4-2{r}^{20}\Lambda-3c_4{r} ^{19}\Lambda+ \left( 6{c_4}^{2}+10c_1 \right) {r}^{16}+ 16\left({c_1}^{2}+2\Lambda c_3 \right) {r}^{14}+21c_4{r}^{15}c_1+54\,{r}^{12}c_3+ 77c_3\left( c_4+\Lambda c_3^{1/2} \right) {r}^{11}\nonumber\\
&+ 96\,c_1\,{r}^{10}c_3+108\,{r}^{9}c_3{}^{3/2}+140\,c_4\,{r} ^{8}c_3{}^{3/2}+165\,c_1\,{r}^{7}c_3{}^{3/2}+96\,{c_3}^{2}{r}^{6}+ 285\,c_3\,{r}^{3}c_3{}^{3/2}+198\,{c_3{}^{3}}\bigg\}\bigg[ \bigg( 3\,c_4\,{r}^{8}+4\,c_1\,{r}^{7}\nonumber\\
&+8\,c_3\,{r}^{3}+11\,c_3{}^{3/2} \bigg) r \left( \Lambda\,{r}^{11}+{r}^{9}+c_4\,{r}^{8}+c_1 \,{r}^{7}+c_3\,{r}^{3}+c_3{}^{3/2} \right) \bigg]^{-1}{dr}\bigg\}\int \! \bigg( \left( c_1+{c_1}^{2}+{c_2}^{2} \right) {r}^{7}10\,c_3\,{r}^{3}+22\,c_3{}^{3/2} \bigg) {r}^{9}\nonumber\\
&+exp\bigg\{-\int \bigg\{4\,{r}^{17}c_4-2\,{r}^{20}\Lambda-3\,c_4\,{r}^{19}\Lambda+ \left( 6\,{c_4}^{2}+10\,c_1 \right) {r}^{16}+21\,c_4\, {r}^{15}c_1+ \left( 16\,{c_1}^{2}+32\,\Lambda\,c_3 \right) {r}^{14}+54{r}^{12}c_3\nonumber\\
&+ \left( 77c_4c_3+ 77\,\Lambda c_3{}^{3/2} \right) {r}^{11}+96c_1\,{r}^{10}c_3+ 108{r}^{9}c_3{}^{3/2}+140c_4\,{r}^{8}c_3{}^{3/2}+165c_1\,{r} ^{7}c_3{}^{3/2}+96\,{c_3}^{2}{r}^{6}+285\,c_3\,{r}^{3}c_3{}^{3/2}+ 198\,{c_3{}^{3}}\bigg\}\nonumber\\
&\bigg[ \left( 3c_4\,{r}^{8}+4c_1\,{r}^{7} +8\,c_3{r}^{3}+11c_3{}^{3/2} \right) r \left( \Lambda {r}^{11}+ {r}^{9}+c_4\,{r}^{8}+c_1\,{r}^{7}+c_3\,{r}^{3}+c_3{}^{3/2} \right) \bigg]^{-1}{dr}\bigg\} \bigg( \Lambda\,{r}^{11}+{r}^{9}+c_4{r}^{8}+c_1\,{r}^{7}\nonumber\\
&+c_3{r}^{3}+c_3{}^{3/2} \bigg) ^{-1}\bigg[ \left( 3\,c_4\,{r}^{8}+4\,c_1\,{r}^{7} +8\,c_3\,{r}^{3}+11\,c_3{}^{3/2} \right) r \left( \Lambda\,{r}^{11}+ {r}^{9}+c_4\,{r}^{8}+c_1\,{r}^{7}+c_3\,{r}^{3}+c_3{}^{3/2} \right) \bigg]^{-1}{dr}\bigg\}\nonumber\\
& \left( 3\,c_4\,{r}^{8}+4\,c_1\,{r}^{7}+8\,c_3\,{r}^{3}+11\,c_3{}^{3/2} \right) ^{-1}{dr}-16\bigg[ -96c_1{r}^{10}c_3{}^{5/2}-12c_4{r}^{26}c_1\Lambda+ \left( -3{c_3}^{2} -{\frac {167}{2}}c_4c_3{}^{3/2}c_1 \right) {r}^{15}\nonumber\\
&- \left( 64\,\Lambda\,c_3{}^{3/2}16\,c_4\,c_3 \right) {r}^{20} - \left( {\frac {199}{2}}c_4\,c_3{}^{3/2}\Lambda+{\frac {79}{4} }{c_4}^{2}c_3+12c_1c_3 \right) {r}^{19}- \left(  \left( 50\,\Lambda\,c_3+5\,{c_1}^{2} \right) c_4+55\,{\Lambda}^{2}c_3{}^{3/2} \right) {r}^{22}\nonumber\\
& - \left( 2{c_1}^ {3}+52c_1c_3\,\Lambda \right) {r}^{21}- \left({\frac { 169}{4}}{c_4}^{2}c_3{}^{3/2}+30c_1\,c_3{}^{3/2} \right) {r}^{ 16}- \left( {c_1}^{2}+34\Lambda c_3+{\frac {15}{4}}{c_4}^{2}c_1 \right) {r}^{23}- \left( \frac{1}2\,{c_4}^{2}+28{\Lambda}^{2}c_3+8{c_1}^{2}\Lambda \right) {r}^{25}\nonumber\\
&- \left( 122c_3c_3{}^{3/2} \Lambda+41\,{c_1}^{2}c_3{}^{3/2}+31c_4\,{c_3}^{2} \right) {r}^{14}- \left( 20{c_1}^{ 2}c_3+37c_4\,c_3{}^{3/2}+40\,\Lambda{c_3}^{2} \right) {r}^{17}- \left( 2c_4\,c_1+\frac{3}4{c_4}^{3} \right) {r}^{24}\nonumber\\
&-\left( {\frac {319}{4}}\,{c_3{}^{3}}\Lambda+{\frac {199} {2}}\,c_4\,c_3{}^{5/2} \right) {r}^{11}- \left( 9/2\,{c_3{}^{3}}+12\,{c_3}^{3} \right) {r}^{9}- \left( 8\,c_1\, \Lambda+{\frac {15}{4}}\,\Lambda\,{c_4}^{2} \right) {r}^{27}+{r}^ {31}{\Lambda}^{2}-{\frac {99}{4}}\,{c_3{}^{9/2}}-6\,{\Lambda}^{2}{r}^ {29}c_1\nonumber\\
&-40\, \left( c_4\,c_3+5/2\,\Lambda\,c_3{}^{3/2} \right) c_1\,{r}^{18}-30\,c_1\,{r}^{13}{c_3}^{2}-12\,c_3\,{r}^{12}c_3{}^{3/2}-51\,{c_3}^{2}{r}^{6}c_3{}^{3/2}-{\frac { 255}{4}}\,c_3\,{r}^{3}{c_3{}^{3}}-{\frac {265}{4}}\,c_4\, {r}^{8}{c_3{}^{3}}-{\frac {255}{4}}c_1{r}^{7}{c_3{}^{3}}\nonumber\\
&-3\,{\Lambda}^{2}{r}^{30}c_4-4\,\Lambda\,c_4\,{r}^{28} \bigg] c_8\,exp\bigg\{\int\bigg\{-2\,{r}^{20}\Lambda-3\,c_4\,{r}^{19}\Lambda+4\,{r}^{17}c_4+ \left( 6\,{c_4}^{2} +10\,c_1 \right) {r}^{16}+21\,c_4\,{r}^{15}c_1+16 \left( {c_1}^{2}+2\Lambda\,c_3 \right) {r}^{14}\nonumber\\
&+54\,{r}^{12} c_3+ \left( 77\,c_4\,c_3+77\,\Lambda\,c_3{}^{3/2} \right) {r}^{11}+96\,c_1\,{r}^{10}c_3+108\,{r}^{9}c_3{}^{3/2}+140\,c_4\,{r}^{8}c_3{}^{3/2}+165c_1{r}^{7}c_3{}^{3/2}+96{c_3} ^{2}{r}^{6}+285c_3{r}^{3}c_3{}^{3/2}\nonumber\\
&+198\,{c_3{}^{3}}\bigg\}\bigg[ \left( 3\,c_4\,{r}^{8}+4\,c_1\,{r}^{7}+8\,c_3\,{r}^{3} +11\,c_3{}^{3/2} \right) r \bigg( \Lambda\,{r}^{11}+{r}^{9}+c_4\,{r} ^{8}+c_1\,{r}^{7}+c_3\,{r}^{3}+c_3{}^{3/2} \bigg) \bigg]^{-1}{dr}\bigg\}-9\, \bigg( {r}^{16}c_4\,\Lambda\nonumber\\
&+{\frac {64}{9}} \,{r}^{11}\Lambda\,c_3+4/9\Lambda \left( 4\,c_1+{c_1}^{2}+{c_2}^{2} \right) {r}^{15}+1/9c_4\, \left( c_1+{c_1}^{2}+{c_2}^{2} \right) {r}^{12}+ \left( {\frac {121}{9}}\Lambda\,c_3{}^{3/2}+{\frac {25}{9}}c_4c_3 \right) {r}^{8}+{ \frac {64}{9}}\,{r}^{5}c_4c_3{}^{3/2}\nonumber\\
&-4/9c_3\left( {c_1}^{2}+{c_2}^{2}-4c_1 \right) {r}^{7}-{\frac {7}{9}}\,c_3{}^{3/2}\, \left( {c_2}^{2}-7\,c_1+{c_1}^{2} \right) {r}^{4}+c_3\,c_3{}^{3/2} \bigg) {r}^{13} \bigg\} \left( 3\,{r}^{29}c_4 +4\,{r}^{28}c_1+8\,{r}^{24}c_3+11\,{r}^{21}c_3{}^{3/2} \right) ^{-1}\,.
\end{align}
The asymptotic form of  $\lambda(r)$ as $r\to \infty$ takes the form:
\begin{align}\label{lam}
&\lambda(r)\approx -{\frac {4 \left( c_1+{c_1}^{2}+{c_2}^{2} \right) \Lambda}{3{\mu}^{4}c_4r}}+{\frac { 16\left( c_1+{c_1}^{2}+{c_2}^{2} \right) \Lambda c_1}{9{\mu}^{4} {c_4}^{2}{r}^{2}}}-{\frac {64 \left( c_1+{c_1}^{2}+{c_2}^{2} \right) \Lambda\,{c_1}^{2}}{27{\mu}^{4} {c_4}^{3}{r}^{3}}}-{\frac {4 \left( c_1+{c_1}^{2}+{c_2}^{2} \right)  \left( -64\Lambda\,{c_1}^{ 3}+27{c_4}^{4} \right) }{81{\mu}^{4}{c_4}^{4}{r}^{4}}}\nonumber\\
&-{\frac {4}{243}}\,{\frac {-27\,{c_1}^{2}{c_4}^{4}-27\,{c_4}^{4}{c_1}^{3}-27\,{c_4}^{4}c_1\,{c_2}^{2}+
810\,{c_4}^{4}\Lambda\,c_3+256\,{c_1}^{5}\Lambda+256\,{
c_1}^{6}\Lambda+256\,{c_2}^{2}\Lambda\,{c_1}^{4}}{{\mu}
^{4}{c_4}^{5}{r}^{5}}}
\,.
\end{align}
Eq.~ (\ref{lam}) shows that when the constant $c_4$ and $\mu$ should not vanish. Moreover, if the constants $c_1$ and $c_2$ which are related to the electric and magnetic fields are vanishing we see that value of the Lagrangian multiplier begins from $O(\frac{1}{r^5})$ which is the order related to the constant $c_3$. This means that in the absence of the electric and magnetic fields, the starting order of the Lagrangian multiplier is  $O(\frac{1}{r^5})$. Now let us analyze the  behavior of the arbitrary function $h$ whose asymptotic as $r\to 0$ takes the form:
\begin{align}\label{has}
h(r)\approx-\frac{3c_4c_5}{2}+c_5r-\frac{c_5r^2}{3c_4}+\frac{2c_5r^3}{27c_4{}^2}-\frac{c_5r^4}{81c_4{}^3}+c_6\equiv C+c_5 r-\frac{c_5r^2}{3c_4}+\frac{2c_5r^3}{27c_4{}^2}-\frac{c_5r^4}{81c_4{}^3}\,.
\end{align}
Eq.~ (\ref{has}) shows that the arbitrary function $h$ has a constant value as $r\to0$.
Finally, the asymptotic form of  $V(r)$  as $r\to \infty$ takes the form:
\begin{align}\label{apot}
V(r)\approx -3\Lambda-\frac{12\Lambda(c_2{}^2+c_1+c_1{}^2)}{rc_4}+\frac{16\Lambda\,c_1(c_1+c_2{}^2+c_1{}^2)}{9r^2c_4{}^2}\,.
\end{align}
Eq.~(\ref{apot}) shows that as $r\to \infty$ we have a constant potential.
In Figure \ref{Fig:1} we show the behavior of the potential ${\tilde V}$, the arbitrary function $h$, and the Lagrangian multiplier $\lambda$.
\begin{figure}
\centering
\subfigure[~The  potential ${\tilde V}$ ]{\label{fig:v}\includegraphics[scale=0.3]{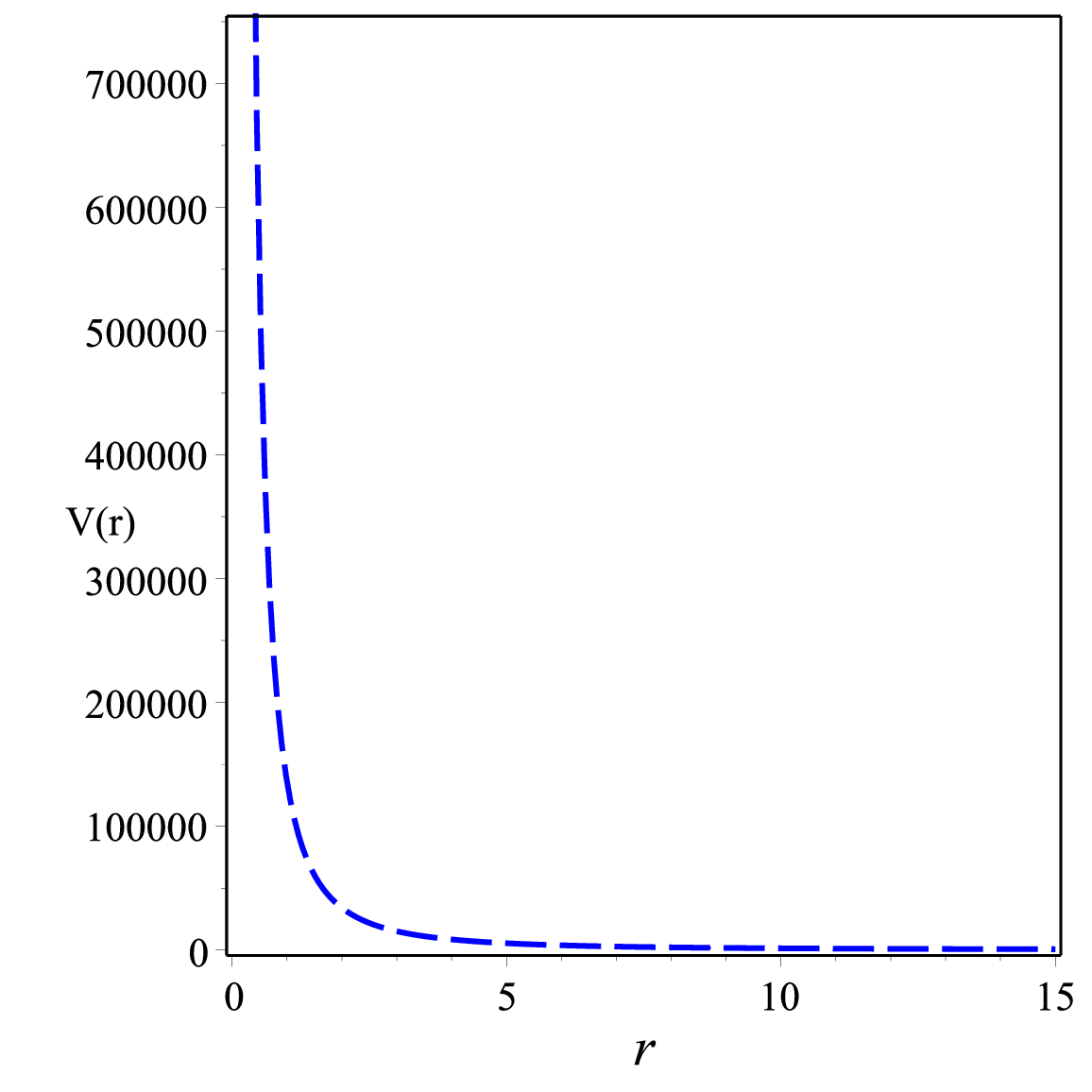}}
\subfigure[~The arbitrary function $h$]{\label{fig:h}\includegraphics[scale=0.3]{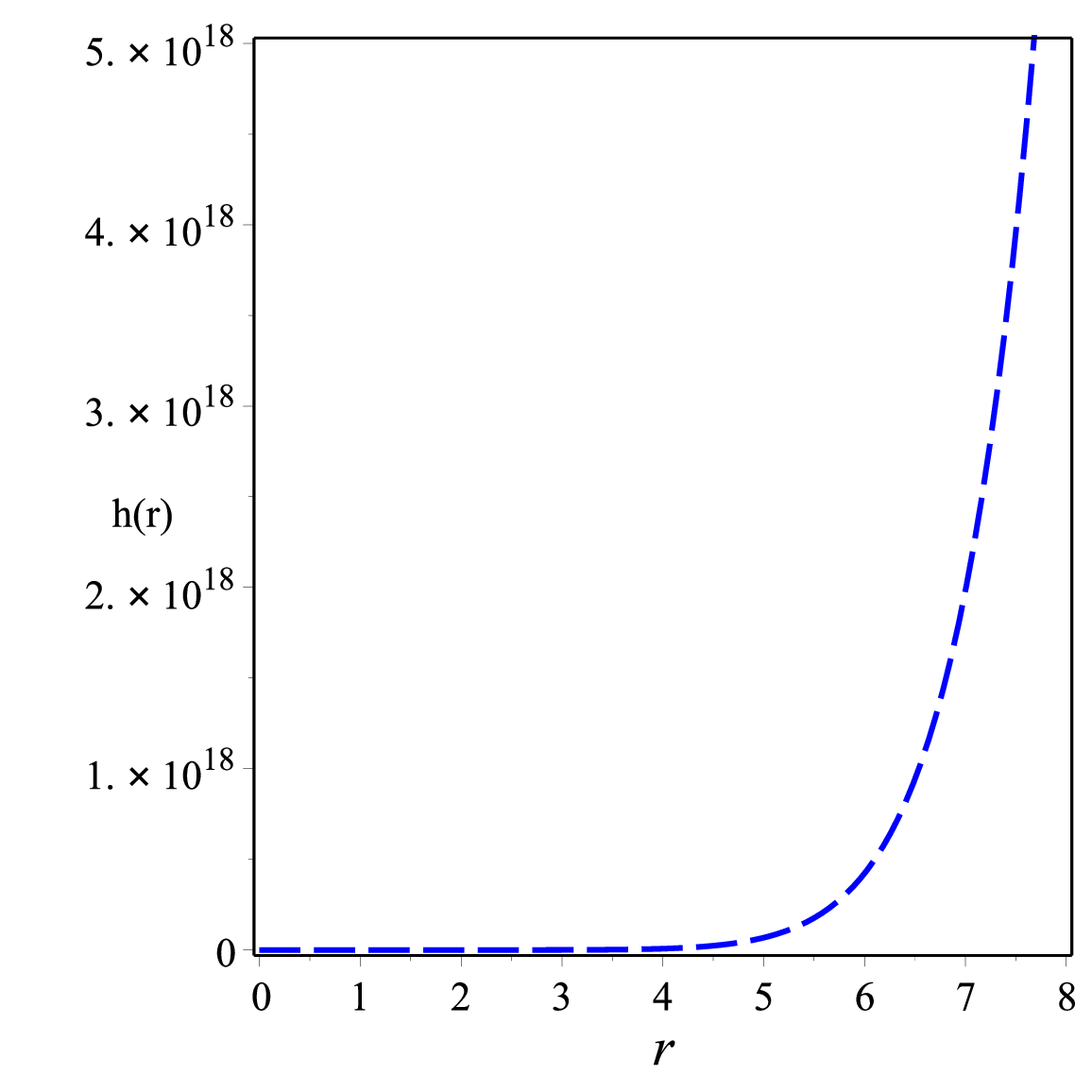}}
\subfigure[~The Lagrangian multiplier $\lambda$]{\label{fig:l}\includegraphics[scale=0.3]{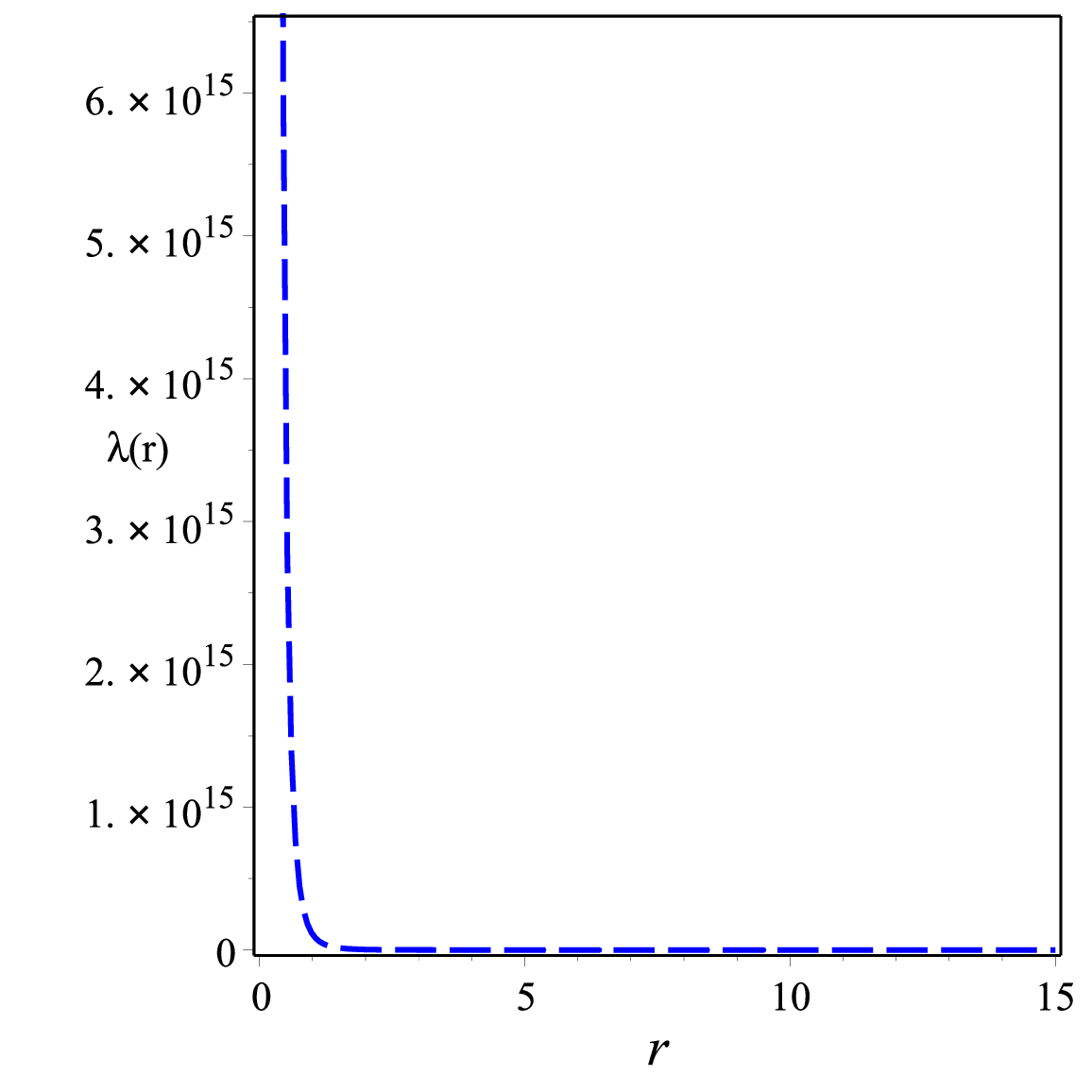}}
\caption[figtopcap]{\small{{Plots of;
Figure~\subref{fig:v} is the potential ${\tilde V}$; Figure~\subref{fig:h} is the behavior of the arbitrary function $h$; Figure~\subref{fig:l} the Lagrangian multiplier. Here we take  the numerical values of the parameters characterized the model as $M=0.01$, $\Lambda=-0.012$,  $q_e=-7$, $q_m=-7$ and $c_3=5$, $c_5=c_6=1$.}}}
\label{Fig:1}
\end{figure} }

\end{document}